# Learning after COVID-19 and the ICT career aspirations: Are students entering the AI era with weaker skills?

Diana Maria Popa, Simona-Vasilica OPREA*, Adela BÂRA
*Department of Economic Informatics and Cybernetics, Bucharest University of Economic Studies, Bucharest, Romania*
*Corresponding author. E-mail: simona.oprea@csie.ase.ro

**Abstract:** This paper examines whether students are entering the generative AI era with sufficiently strong educational foundations, focusing on the relationship between learning environments and changes in ICT-related career aspirations across countries. The analysis uses country-level data from PISA 2018 and 2022, combining indicators of student autonomy, digital skills and teacher support. A mixed-method approach is applied, including descriptive statistics, regression analysis, clustering, latent representation learning (using Variational Autoencoder-VAE), discriminant analysis and probabilistic modeling to capture both observable and latent dimensions of educational readiness. Unlike prior research that treats learning loss, digital skills and career expectations separately, our analysis integrates them within a comparative longitudinal framework. It shifts the focus from short-term post-pandemic effects to the structural capacity of education systems to prepare students for digital and AI-driven labor markets. Results show a global but uneven increase in ICT career aspirations. Digital skills emerge as the strongest and most consistent predictor, while teacher support plays a complementary role. Autonomy shows weaker, context-dependent effects. Educational readiness is multidimensional, and ICT aspirations evolve relatively independently from other career domains.

## 1. Introduction

The expansion of digital technologies, intensified by the COVID-19 pandemic and further accelerated by the emergence of gen-AI, has transformed both the conditions of learning and the skills required for future labor market participation. In this context, a key question is whether students are entering the gen-AI era with sufficiently strong educational foundations to support both academic performance and future-oriented digital career aspirations. Although recent research has extensively examined post-pandemic learning loss, digital inequalities and the growing relevance of ICT- and AI-related skills, these studies have often remained disconnected. Further, studies on educational outcomes tend to focus on performance decline, while labor market research emphasizes the rise of digital occupations without systematically linking these developments to the structural conditions of education systems. As a result, limited attention has been paid to the relationship between learning environments, digital preparedness and changes in students' occupational expectations across countries.

Thus, our study addresses this gap by investigating whether structural characteristics of learning environments are associated with changes in ICT-related career aspirations. The analysis is based on country-level indicators derived from PISA 2018 and PISA 2022 [1], [2]. Measures of student autonomy, digital learning skills and perceived teacher support were operationalized from variables and indices available in the PISA database and aggregated at country level for cross-national comparison.

The main objective of our study is to examine whether education systems characterized by stronger autonomy, more developed digital skills and higher levels of teacher support are also those in which ICT-related career aspirations have grown more substantially between 2018 and 2022. A second objective is to assess whether these relationships are consistent across Mathematics, Reading and Science. A third objective is to identify broader structural profiles of educational readiness and evaluate whether such profiles help explain cross-national heterogeneity in ICT aspiration dynamics. These objectives are reflected in the following research questions:

*RQ1:* How do structural characteristics of learning environments (operationalized through student autonomy, digital learning skills and perceived teacher support) relate to cross-national variation in academic outcomes and ICT-related career aspirations?

*RQ2:* To what extent do the effects of autonomy and digital skills exhibit cross-domain consistency across Mathematics, Reading and Science, and what does this reveal about the generality of digital learning conditions across education systems?

*RQ3:* Can latent and observable configurations of educational readiness for the digital and AI era account for longitudinal changes in students' ICT-related occupational expectations between 2018 and 2022?

The originality of the study lies in its integrated perspective. Rather than examining post-pandemic learning outcomes, digital skills or career expectations in isolation, our study connects them within a comparative longitudinal framework. It thus shifts the focus from short-term academic disruption to the broader question of whether education systems are preparing students for participation in increasingly digital and AI-mediated societies. Methodologically, it combines descriptive analysis, regression-based inference, clustering, latent representation learning, discriminant classification and probabilistic modeling in order to capture both observable and latent dimensions of educational readiness. By linking educational structures to changes in ICT-related aspirations, our study contributes to the literature on comparative education, digital transformation and future skills. It also studies that the transition to the generative AI era should be understood not only as a technological shift, but also as a structural educational challenge shaped by institutional, pedagogical and socio-economic conditions.

## 2. Literature review

The COVID-19 pandemic and the expansion of digital technologies have significantly transformed education systems, reshaping both learning environments and students' skill development. An increasing body of empirical research documents substantial learning losses and increasing educational inequalities across countries. Evidence shows that school closures led to declines in academic performance, particularly in domains such as mathematics and science [1], [2], [3], [4], [5]. Systematic reviews further confirm that learning loss represents a global phenomenon, with heterogeneous effects depending on socio-economic context and access to resources [6], [7].

Beyond performance decline, the pandemic has amplified pre-existing inequalities in education systems. Several studies emphasize that disadvantaged students were disproportionately affected due to limited access to digital infrastructure and learning support [8]. These findings are supported by international reports highlighting long-term risks for human capital development and educational equity [9].

At the same time, the increasing integration of digital technologies has shifted attention from access to effective use, giving rise to the concept of the "second digital divide". Early research shows that inequalities in digital skills and usage significantly affect students' ability to navigate online information and learning environments [10], [11]. More recent cross-national studies based on PISA data also confirm that digital competencies are strongly associated with academic achievement and educational performance [12], [13], [14], [15], [16], [17], [18], [19].

In parallel, the role of artificial intelligence (AI) and digital tools in education has gained more importance. Research indicates that AI-enabled learning environments can enhance knowledge acquisition, assessment processes and student engagement, while also introducing new pedagogical challenges [20], [21], [22]. Empirical evidence further suggests that digital and AI-based applications can support learning beyond traditional classroom settings, although their effectiveness depends on institutional and contextual factors [23]. At the system level, international assessments highlight the importance of digital readiness and learning environments in shaping educational outcomes [24], [25].

Another important strand of literature focuses on the relationship between digital learning and career aspirations, particularly in ICT-related fields. Studies show that exposure to ICT and the development of digital skills influence students' attitudes toward technology-oriented careers and their professional expectations [26], [27], [28], [29]. Research on higher education students further indicates that career aspirations in IT fields are shaped by both skill development and educational experiences [30], [31].

More recent contributions highlight that digital career aspirations are influenced by broader socio-economic and contextual factors. Evidence suggests that gender, access to digital resources and learning opportunities play an important role in shaping students' orientation toward digital professions [32], [33], indicating that ICT-related career expectations are not only determined by educational outcomes, but also by structural and contextual conditions.

These studies suggest that learning environments, digital competencies and career aspirations are closely interconnected, yet often analyzed separately. While existing research provides important insights into learning loss, digital skills and ICT-related career orientation, several gaps remain. First, most studies examine these dimensions independently rather than within an integrated framework. Second, there is limited cross-national evidence linking structural learning conditions to changes in career aspirations over time. Third, few studies explicitly address the transition toward the generative AI era as a systemic educational challenge. Therefore, our study addresses these gaps by integrating cross-national PISA data with indicators of student autonomy, digital skills and teacher support, and by linking these structural characteristics to changes in ICT-related career aspirations between 2018 and 2022. By combining these dimensions within a unified analytical framework, the research contributes to a deeper understanding of educational readiness in the context of digital and AI transformation.

## 3. Methodology

### 3.1 Input data

The empirical foundation of our study is based on a harmonized cross-national dataset constructed from multiple publicly available sources related to educational systems, student learning environments and career aspirations in the context of digital transformation. The core of the analysis relies on microdata and derived indicators from the OECD Programme for International Student Assessment (PISA) waves 2018 and 2022, complemented by a set of synthetic indicators designed to capture structural readiness for the digital and AI era. All data sources are open-access and internationally comparable, ensuring transparency and reproducibility of the empirical analysis. The countries included in the final analytical sample are those for which complete information is available across both PISA waves and for all constructed indicators, as summarized in Table 1.

Table 1. Input data sources

| No. | Data source (OECD) | Description | Years | Domains | Variables used |
|---|---|---|---|---|---|
| **1** | Table II.B1.2.4 | Change between 2018 and 2022 in students' career expectations | 2018, 2022 | Career expectations | delta_ict, delta_health, delta_sci_eng, delta_sci_tech |
| **2** | Table II.B1.2.26 | Experience with learning at home and mathematics performance | 2022 | Mathematics | autonomy_effect, digital_effect, teacher_support_effect |
| **3** | Table II.B1.2.27 | Experience with learning at home and science performance | 2022 | Science | autonomy_effect, digital_effect, teacher_support_effect |
| **4** | Table II.B1.2.28 | Experience with learning at home and reading performance | 2022 | Reading | autonomy_effect, digital_effect, teacher_support_effect |

The dataset integrates information on students' learning environments, psychosocial characteristics and occupational expectations, thereby enabling a longitudinal perspective on how career aspirations evolve in relation to changes in educational conditions. From *PISA 2022*, indices capturing *student autonomy*, *digital skills* and perceived *teacher support* were extracted and aggregated at the country level in order to facilitate cross-system comparisons.

From *PISA 2018* and *2022*, indicators reflecting students' interest in ICT-related careers and in science and engineering occupations were derived, and *intertemporal changes* between the two waves were computed. In parallel, system-level contextual information reflecting digital preparedness was incorporated in the form of a composite *AI Readiness Index*. The indicators used in this study were operationalized from existing data sources. More specifically, the measures of student autonomy, digital skills and perceived teacher support were derived from variables and indices available in the PISA 2022 database, then aggregated at country level to enable cross-national comparison. Likewise, indicators of *interest in ICT-related careers* and in *science* and *engineering* occupations were derived from PISA 2018 and PISA 2022, and their intertemporal changes were computed comparatively across the two waves.

The study's added methodological contribution therefore lies not in creating completely new indicators, but in selecting, harmonizing, aggregating and combining established measures in a comparative longitudinal framework. In addition, the AI Readiness Index was included as an external composite contextual indicator, rather than being constructed within the present study.

Given the heterogeneity of the original data sources and measurement scales, a preprocessing pipeline was implemented to harmonize all inputs into a unified analytical framework. Country identifiers were aligned across PISA waves, variables were standardized to ensure cross-country comparability and intertemporal differences were calculated in order to capture dynamic changes rather than static levels. This harmonization process resulted in a balanced panel of education systems observed over the 2018–2022 period.

To capture latent profiles of educational preparedness for the digital and AI era, a Variational Autoencoder (VAE) was employed. The model was trained on a multidimensional set of standardized country-level indicators describing digital infrastructure, educational resources and learning environment characteristics. The VAE projects these high-dimensional inputs into a low-dimensional latent space, which is interpreted as a continuous representation of structural readiness for digital and AI-related transformations. This latent representation enables both the visualization of similarities between education systems and subsequent statistical association analyses with changes in students' ICT-related career aspirations.

The analytical strategy also combines descriptive exploration with regression-based inference. Associations between student autonomy, digital skills and learning outcomes were estimated while controlling for *socio-economic background*, following the PISA methodological framework. Cross-domain consistency was examined by comparing the strength of these relationships across Mathematics, Reading and Science. Furthermore, interaction effects between autonomy, digital skills and teacher support were explored in order to assess whether supportive learning environments amplify or attenuate the benefits of increased student autonomy and self-directed engagement in digital learning contexts. At the system level, the relationship between the AI Readiness Index and changes in ICT-related career interest was examined to evaluate whether structural preparedness is reflected in students' occupational expectations.

### 3.2 Framework

The analytical framework is built as a fully reproducible, end-to-end quantitative pipeline that transforms raw international education indicators into structured analytical representations used for descriptive, inferential and representation learning analyses. The framework is explicitly grounded in the computational workflow implemented in Python and reflects the exact sequence of transformations applied to the data, from extraction and harmonization of PISA tables to the estimation of latent structural profiles and predictive models of ICT career aspiration dynamics.

The pipeline follows a deterministic flow of operations: raw tabular indicators are first parsed and cleaned, country-level analytical variables are constructed through rule-based extraction procedures, multivariate structures are explored through clustering and dimensionality reduction, and the resulting representations are linked to longitudinal changes in occupational expectations using regression, classification and probabilistic modeling.

The following notation is used throughout the analytical framework. Let $i = 1, \dots, N$ denote the index of education systems (countries). The main variables are defined as follows:

- $A_i$-effect of student autonomy on performance;
- $D_i$-effect of digital learning skills on performance;
- $T_i$-effect of teacher support on performance.

The vector of learning environment characteristics is defined as: $X_i = (A_i, D_i, T_i)$. Changes in career aspirations are denoted as: ΔICT$i$, ΔHealth$i$, ΔSciEng$i$, ΔSciTech$i$. Domain-specific variables are indexed by $d \in$ {math,reading,science}: $X_i^{(d)} = (A_i^{(d)}, D_i^{(d)}, T_i^{(d)})$. Discretized variables are denoted using a tilde: $\tilde{X}_i$. Latent representations obtained from the variational autoencoder are denoted as: $z_i = (z_{i1}, z_{i2})$. The AI Readiness Index is defined as: $AI_i$.

#### 3.2.1 Data extraction, harmonization and numerical encoding

For each education system $i$, the framework extracts three core learning environment indicators derived from PISA 2022: (i) the effect of student autonomy in learning on academic performance, (ii) the effect of digital learning skills on academic performance and (iii) the effect of teacher availability for support on academic performance. In the source tables, these effects are reported as score differences after accounting for students' and schools' socio-economic profiles and are encoded as string values that may contain non-numeric characters (e.g., confidence intervals, symbols or missing-value labels such as “m”).

To convert these heterogeneous entries into numerical values suitable for quantitative analysis, a rule-based parsing function is applied to each table cell. Let $s_i$ denote the raw string associated with country $i$. The numerical encoding is defined as:

$$x_i = \begin{cases} \text{numeric}(s_i), & \text{if } s_i \notin \{\text{missing}, \text{“m”}\} \\ \text{NA}, & \text{otherwise} \end{cases} \quad (1)$$

where $\text{numeric}(s_i)$ denotes the extraction of the signed integer or floating-point component from the string using regular expression matching. This operation ensures that only valid quantitative score differences are retained, while non-informative or missing entries are systematically filtered out. In addition to numerical parsing, country identifiers are normalized by removing annotation symbols (such as asterisks or footnote markers) present in the OECD tables.

In parallel, longitudinal indicators of occupational expectations are constructed using PISA data on students' career aspirations in 2018 and 2022. For each country $i$, percentage changes are extracted for four professional domains: ICT-related occupations, health professions, science and engineering occupations and science-related technical occupations. The temporal variation in ICT career interest is computed as:

$$\Delta\text{ICT}_i = \text{ICT}_{i,2022} - \text{ICT}_{i,2018} \quad (3)$$

where $\text{ICT}_{i,t}$ denotes the percentage of students in country $i$ expressing interest in ICT professions in year $t$. Analogous change indicators are computed for the remaining occupational domains and used in comparative and multivariate analyses.

After numerical encoding and harmonization, all indicators are merged into a unified country-level analytical matrix:

$$X = \begin{bmatrix} A_1 & D_1 & T_1 & \Delta\text{ICT}_1 & \Delta\text{Health}_1 & \Delta\text{SciEng}_1 & \Delta\text{SciTech}_1 \\ A_2 & D_2 & T_2 & \Delta\text{ICT}_2 & \Delta\text{Health}_2 & \Delta\text{SciEng}_2 & \Delta\text{SciTech}_2 \\ \vdots & \vdots & \vdots & \vdots & \vdots & \vdots & \vdots \\ A_n & D_n & T_n & \Delta\text{ICT}_n & \Delta\text{Health}_n & \Delta\text{SciEng}_n & \Delta\text{SciTech}_n \end{bmatrix} \quad (2)$$

where $A_i$, $D_i$ and $T_i$ denote the autonomy, digital skills and teacher support effects on performance for country $i$.

### 3.2.2 Multivariate profiling of learning environments

After harmonization, the three learning environment indicators, student autonomy in learning, digital learning skills and teacher support availability, are transformed through standardization in order to remove scale effects and ensure comparability across dimensions. For each indicator $x \in \{A, D, T\}$, a z-score transformation is applied:

$$z_{i,x} = \frac{x_i - \mu_x}{\sigma_x} \quad (4)$$

where $\mu_x$ and $\sigma_x$ denote the mean and standard deviation of indicator $x$ across all countries, ensuring that each dimension contributes equally to distance-based methods and prevents any single indicator from dominating the clustering solution due to scale differences.

Each education system $i$ is thus represented as a point in a three-dimensional standardized feature space:

$$Z_i = \left(A_i^{(z)}, D_i^{(z)}, T_i^{(z)}\right) \quad (5)$$

which defines a joint configuration of learning environment characteristics.

To identify recurrent structural profiles of education systems, unsupervised clustering is applied to the standardized feature vectors using the KMeans algorithm. The objective of KMeans is to partition the set of countries into $K$ clusters $\{C_1, \dots, C_K\}$, such that the within-cluster sum of squared deviations is minimized:

$$J = \sum_{k=1}^{K} \sum_{i \in C_k} | Z_i - \mu_k |^2 \quad (6)$$

where $\mu_k$ denotes the centroid of cluster $C_k$, computed as the mean of the standardized feature vectors assigned to that cluster. The optimization proceeds iteratively by alternating between cluster assignment and centroid update until convergence.

The resulting clusters are interpreted as typologies of education systems characterized by distinct configurations of autonomy, digital preparedness and teacher support. For example, clusters may correspond to systems combining high autonomy with strong digital skills but moderate teacher support or conversely to systems with strong teacher support but weaker digital preparedness. These typologies provide an intermediate abstraction layer between raw country-level indicators and higher-level latent representations derived in subsequent stages of the pipeline.

By constructing these multivariate profiles, the framework enables a structured comparison of education systems beyond univariate rankings, capturing interaction patterns between learning environment dimensions, constituting an important link between descriptive analysis and the downstream modeling components, as the identified clusters are later related to variations in students' ICT-related career aspirations and to latent representations obtained through representation learning models.

**3.2.3 Cross-domain consistency analysis**

To assess the robustness and generalizability of the extracted learning environment effects, the analytical pipeline extends the indicator construction procedure to multiple cognitive domains, namely Mathematics, Reading and Science. For each domain $d \in \{\text{math}, \text{reading}, \text{science}\}$, domain-specific effects of student autonomy, digital learning skills and teacher support are extracted using the same harmonization and parsing procedure applied in the baseline analysis.

Formally, for each country $i$ and domain $d$, the learning environment configuration is represented as:

$$\mathrm{X}_i^{(d)} = \left(A_i^{(d)}, D_i^{(d)}, T_i^{(d)}\right) \tag{7}$$

where $A_i^{(d)}$, $D_i^{(d)}$and $T_i^{(d)}$denote the estimated performance score differences associated with autonomy, digital skills and teacher support in domain $d$, after accounting for students' and schools' socio-economic background, as reported in the PISA 2022 data.

Cross-domain consistency is evaluated by computing pairwise Pearson correlation coefficients between homologous indicators across domains. For instance, the stability of autonomy effects between Mathematics and Reading is assessed as:

$$\rho\left(A^{\text{math}}, A^{\text{reading}}\right) = \text{corr}\left(A_i^{\text{math}}, A_i^{\text{reading}}\right) \tag{8}$$

and analogously for Science. High correlation values indicate that countries exhibiting strong autonomy-related performance effects in one domain tend to exhibit similar patterns in other domains, suggesting that the learning environment operates as a system-level structural feature rather than as a subject-specific phenomenon. The same procedure is applied to digital learning effects and teacher support effects, yielding cross-domain correlation matrices that quantify the degree of structural coherence of learning environments across subjects. Beyond correlational consistency, domain-specific differences in effect magnitudes are formally tested using paired statistical tests. In particular, paired $t$-tests are applied to digital learning effects across domains:

$$H_0\colon \mathbb{E}[D_i^{(math)}] = \mathbb{E}\left[D_i^{(reading)}\right] H_0\colon \mathbb{E}[D_i^{(math)}] = \mathbb{E}[D_i^{(science)}]\ H_0\colon \mathbb{E}[D_i^{(reading)}] = \mathbb{E}[D_i^{(science)}] \tag{9}$$

where the null hypotheses state that the mean digital learning effect does not differ across cognitive domains. Rejecting these hypotheses provides evidence that the strength of digital learning benefits is domain-dependent, whereas non-rejection supports the interpretation of digital preparedness as a transversal learning environment factor.

In addition to statistical testing, the joint configuration of autonomy and digital learning effects across domains is examined through bivariate projections in a shared coordinate system. Overlaying the domain-specific point clouds allows for a direct visual comparison of how education systems reposition themselves when moving from Mathematics to Reading or Science. The geometric comparison complements the correlation-based evidence by revealing whether cross-domain differences arise from systematic shifts or from isolated outliers, validating whether the estimated relationships are structurally stable properties of education systems or whether they are contingent on the specific cognitive domain under consideration. Thus, the framework strengthens the interpretation of autonomy, digital skills and teacher support as systemic features of educational preparedness rather than domain-specific artifacts.

**3.2.4 Learning latent structural representations via VAE**

To capture non-linear dependencies and latent structural regularities that are not directly observable in the original indicator space, the analytical framework employs a Variational Autoencoder (VAE). While linear projections and clustering methods capture only shallow geometric relationships between autonomy, digital skills and teacher support, VAE enables the discovery of non-linear manifolds that encode deeper structural similarities between education systems. Each education system $i$ is represented by a standardized feature vector:

$$\mathrm{X}_i = (A_i, D_i, T_i) \tag{10}$$

and these vectors constitute the input layer for VAE.

The encoder network implements a non-linear mapping from the observed space to a low-dimensional latent distribution. For each input $\mathrm{X}_i$, the encoder produces the parameters of a multivariate Gaussian posterior:

$$q_\phi(\mathrm{z}_i \mid \mathrm{X}_i) = \mathcal{N}\left(\mu_i, \mathrm{diag}\left(\sigma_i^2\right)\right) \tag{11}$$

where $\mu_i \in \mathbb{R}^2$ and $\sigma_i^2 \in \mathbb{R}^2$ denote the learned mean and variance vectors of the two-dimensional latent representation. Latent samples are drawn using:

$$\mathrm{z}_i = \mu_i + \sigma_i \odot \varepsilon, \varepsilon \sim \mathcal{N}(0, \mathrm{I}) \tag{12}$$

which ensures differentiability of the stochastic sampling process and enables end-to-end optimization via backpropagation. The decoder network implements the inverse mapping, reconstructing the original input vector from the latent code:

$$\widehat{\mathrm{X}}_i = f_\theta(\mathrm{z}_i) \tag{13}$$

where $f_\theta(\cdot)$ denotes a non-linear neural mapping parameterized by $\theta$. VAE is trained by minimizing a regularized variational objective that balances reconstruction fidelity with latent space regularization:

$$\mathcal{L}(\mathrm{X}_i) = \| \mathrm{X}_i - \widehat{\mathrm{X}}_i \|^2 + \mathrm{KL}\left(q_\phi(\mathrm{z}_i \mid \mathrm{X}_i) \parallel p(\mathrm{z})\right) \tag{14}$$

where $p(\mathrm{z}) = \mathcal{N}(0, \mathrm{I})$ is the isotropic Gaussian prior and the Kullback–Leibler divergence term enforces smoothness and regularity in the latent space. The first term ensures faithful reconstruction of the observed learning environment indicators, while the second term constrains the learned representations to occupy a continuous and structured latent manifold.

The resulting two-dimensional latent space provides a continuous embedding of education systems, in which geometric proximity reflects similarity in their joint learning environment configurations. Countries that share similar combinations of autonomy, digital preparedness and teacher support are mapped to neighboring regions of the latent manifold, while structurally divergent systems occupy distant regions.

The latent representation is interpreted as a structural readiness space for digital and AI-related transformations in education. To operationalize its representation in subsequent analyses, a composite *AI Readiness Index* is constructed as a linear aggregation of the latent coordinates:

$$\mathrm{AI}_i = \alpha z_{i1} + (1 - \alpha) z_{i2} \tag{15}$$

with $z_{i1}$ and $z_{i2}$ denoting the two latent dimensions and $\alpha \in [0,1]$ is set to 0.5 in the empirical implementation. The composite index provides a parsimonious scalar proxy for each education system's latent positioning within the structural readiness space and is subsequently used in correlation and regression analyses linking educational preparedness to changes in ICT-related career aspirations.

### 3.2.5 Predictive modeling of ICT career aspiration dynamics

The final stage of the analytical framework establishes an explicit empirical link between structural learning environment indicators, latent representations of educational readiness and longitudinal changes in students' ICT-related career aspirations between 2018 and 2022. This step operationalizes the outcome dimension of the pipeline and translates descriptive and latent representations into inferential and predictive models. First, a multivariate linear regression model is estimated in order to quantify marginal associations between learning environment indicators and the observed change in ICT career expectations. For each education system $i$, the change in ICT aspirations is modeled as:

$$\Delta\mathrm{ICT}_i = \beta_0 + \beta_1 A_i + \beta_2 D_i + \beta_3 T_i + \varepsilon_i \tag{16}$$

where $\varepsilon_i$ is an idiosyncratic error term. The regression coefficients $\beta_1, \beta_2, \beta_3$ provide direct estimates of the partial association between each learning environment dimension and changes in ICT-related career interest, conditional on the remaining indicators. This specification corresponds to the empirical implementation in which standardized predictors are used to ensure comparability of effect magnitudes across dimensions.

Second, a classification-based perspective is introduced through Linear Discriminant Analysis (LDA), which models ICT career dynamics as a categorical outcome. Education systems are dichotomized into groups characterized by high versus low growth in ICT aspirations based on a threshold applied to $\Delta\mathrm{ICT}_i$. The discriminant function is defined as:

$$\delta(\mathrm{X}_i) = \mathrm{w}^\top \mathrm{X}_i \tag{17}$$

where w denotes the vector of discriminant weights estimated via maximum likelihood. This formulation identifies the linear combination of autonomy, digital skills and teacher support that best separates education systems exhibiting above-median growth in ICT aspirations from those exhibiting stagnation or decline. Model performance is evaluated using stratified $K$-fold cross-validation, ensuring that reported classification accuracy reflects out-of-sample predictive performance rather than in-sample fit. The relative magnitude and sign of the discriminant coefficients further provide an interpretable ranking of the predictors' contribution to ICT aspiration growth.

Third, the framework incorporates counterfactual simulations in order to move beyond purely associative interpretations and to provide an interpretable approximation of potential intervention effects. Using the fitted linear regression model, standardized autonomy levels are perturbed by one standard deviation while holding digital skills and teacher support constant:

$$A_i^{\mathrm{cf}} = A_i + 1 \tag{18}$$

Predicted changes in ICT aspirations under this counterfactual scenario are then computed as:

$$\widehat{\Delta\mathrm{ICT}}_i^{\mathrm{cf}} = \hat{\beta}_0 + \hat{\beta}_1 A_i^{\mathrm{cf}} + \hat{\beta}_2 D_i + \hat{\beta}_3 T_i \tag{19}$$

The difference between $\widehat{\Delta\mathrm{ICT}}_i^{\mathrm{cf}}$ and the baseline prediction provides an estimate of the marginal impact associated with a one-standard-deviation increase in student autonomy. Although this procedure does not constitute a causal identification strategy, it offers a transparent and policy-relevant quantification of the leverage that learning environment reforms related to autonomy could exert on ICT career orientation dynamics under the maintained linear model assumptions.

Finally, latent readiness representations obtained from VAE are integrated into this stage by correlating the composite AI Readiness Index with $\Delta\mathrm{ICT}_i$ and by visualizing ICT aspiration changes across the latent embedding space, enabling the interpretation of ICT career dynamics in terms of isolated learning environment indicators and also in relation to the broader structural configuration of education systems captured by the non-linear latent manifold.

**3.2.6 Probabilistic modeling of cross-domain career aspiration dependencies**

To capture potential dependency structures between changes in ICT-related career aspirations and contemporaneous shifts in other occupational domains, the framework incorporates a probabilistic graphical modeling component based on Bayesian networks. For each education system $i$, the vector of longitudinal aspiration changes is defined as:

$$\mathrm{Y}_i = (\Delta\mathrm{Health}_i, \Delta\mathrm{SciEng}_i, \Delta\mathrm{SciTech}_i, \Delta\mathrm{ICT}_i) \tag{20}$$

where each component represents the percentage-point change in the share of students expecting a career in the corresponding occupational domain between PISA 2018 and PISA 2022.

Given the limited sample size at country level and the continuous nature of the indicators, all change variables are discretized into ordinal categories (low, medium, high) using quantile-based binning, ensuring that the joint distribution of aspiration dynamics can be modeled within a discrete Bayesian network framework, while preserving relative rankings across education systems. Formally, each continuous variable $\Delta X_i$ is mapped to a categorical variable $\tilde{X}_i \in \{\mathrm{low}, \mathrm{medium}, \mathrm{high}\}$ according to its empirical quantiles.

The network structure is learned directly from data using a score-based hill-climbing algorithm. Let $G$ denote a candidate directed acyclic graph (DAG) over the nodes corresponding to the four occupational change variables. Structure learning aims to identify the graph $G^*$ that maximizes a penalized likelihood score:

$$G^* = \arg\max_{G \in \mathcal{G}} \mathrm{Score}(\, G \mid \widetilde{\mathrm{Y}} \,) \quad (21)$$

where $\mathcal{G}$ denotes the space of admissible DAGs and the scoring function is instantiated using the Bayesian Information Criterion (BIC), balancing model fit and structural complexity. This procedure operationalizes the assumption that interdependencies between occupational aspiration changes can be approximated by a sparse directed graphical structure.

Given the learned structure $G^*$, the parameters of the Bayesian network are estimated using maximum likelihood. For each node $Y_j$ with parent set $\mathrm{Pa}(Y_j)$, the conditional probability tables are estimated as:

$$\hat{P}(\, Y_j = y \mid \mathrm{Pa}(Y_j) = \pi \,) = \frac{N(Y_j = y, \mathrm{Pa}(Y_j) = \pi)}{N(\mathrm{Pa}(Y_j) = \pi)} \quad (22)$$

where $N(\cdot)$ denotes empirical frequency counts in the discretized dataset.

The resulting Bayesian network provides a compact representation of the joint distribution:

$$P(\widetilde{\mathrm{Y}}_i) = \prod_j P(\, \tilde{Y}_{ij} \mid \mathrm{Pa}(\tilde{Y}_{ij})) \quad (23)$$

allowing for conditional inference and scenario analysis. In particular, probabilistic queries of the form:

$$P(\, \Delta\mathrm{ICT}_i = y \mid \Delta\mathrm{SciEng}_i = \mathrm{low}\,) \quad (24)$$

are computed using exact inference (variable elimination) in order to assess whether changes in ICT aspirations are conditionally associated with concurrent dynamics in science and engineering aspirations. These queries enable the exploration of substitution and complementarity patterns between occupational domains, such as whether declines in STEM-related aspirations are probabilistically linked to stagnation or growth in ICT-related career interest. Importantly, the Bayesian network is interpreted as a descriptive probabilistic model rather than a causal structure. The methodological flowchart is presented in Figure 1.

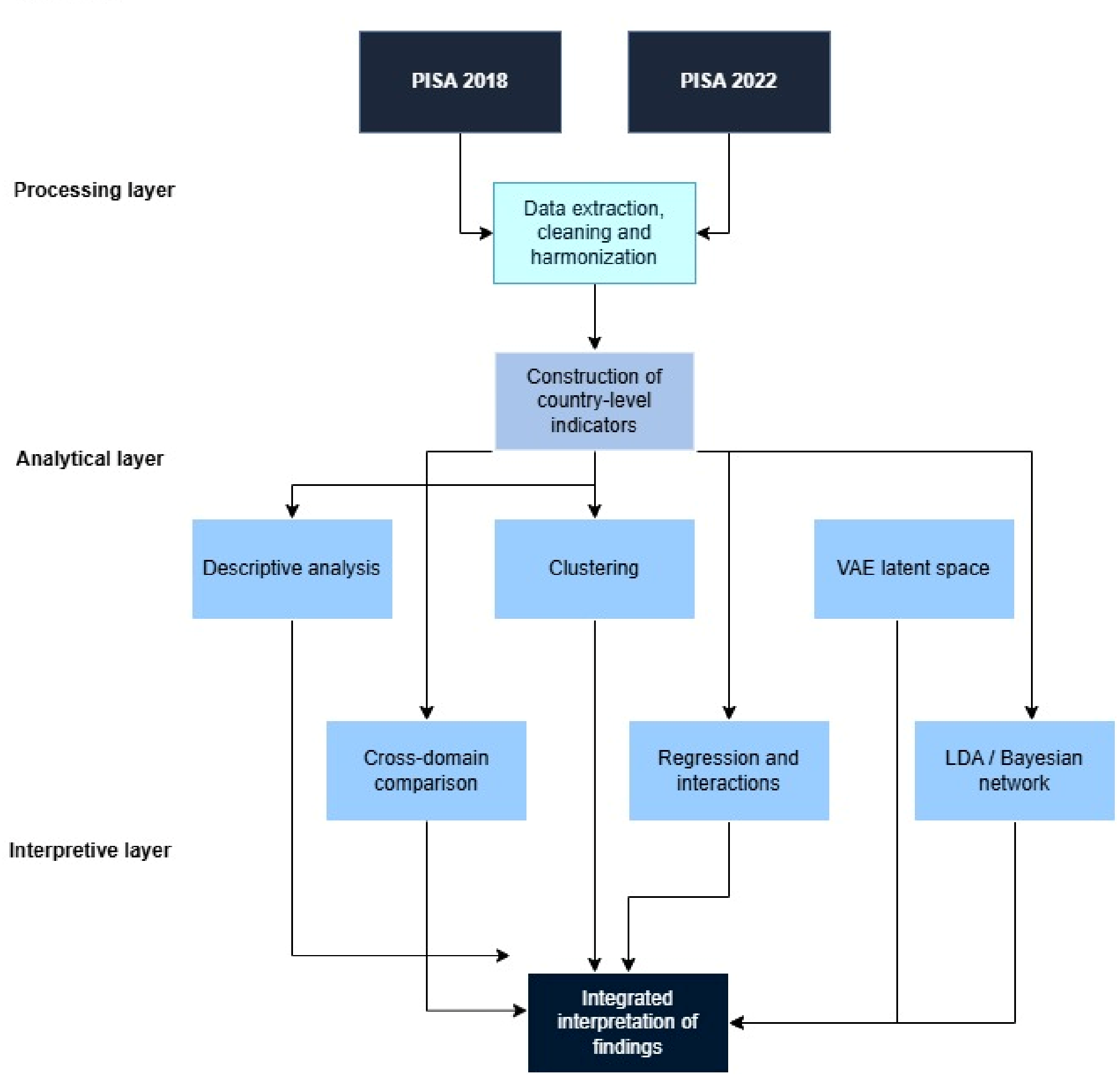

Figure 1. Methodological workflow of the study

## 4. Results

This section presents the results obtained from the application of the analytical framework to the harmonized PISA 2018–2022 datasets. The analysis combines descriptive exploration, latent profiling and multivariate modeling in order to examine the relationship between learning environments and changes in ICT-related career aspirations.

### 4.1. Changes in career aspirations between 2018 and 2022

The first layer of analysis examines the distribution and cross-domain structure of changes in students' career expectations between PISA 2018 and PISA 2022 (Figure 2). The focus is placed on ICT-related occupations, complemented by comparative evidence from health professions, science and engineering occupations and science-related technical careers, establishing the empirical baseline for the subsequent modeling stages.

The distribution of country-level changes in students' interest in ICT-related careers between 2018 and 2022 is positively skewed, with most observations concentrated between 0 and 2 percentage points, indicating that most education systems experienced moderate increases in ICT career aspirations. The presence of a pronounced right tail, with several countries exhibiting large positive changes ($\Delta ICT > 4$), points to the existence of outliers characterized by accelerated growth in ICT-related career interest, reflecting differences in national contexts, including digital transformation dynamics and the perceived attractiveness of ICT-related occupations. At the same time, a small number of negative values indicate that not all countries experienced growth in ICT aspirations, suggesting that structural constraints or alternative career orientations may still shape students' expectations in certain education systems. Figure 3 ranks the ten countries with the largest positive changes in ICT career interest. The large positive deltas observed for countries such as Turkey, Kosovo, Saudi Arabia and Uruguay indicate a pronounced reorientation of student aspirations toward ICT professions, showing that, in some education systems, the perceived attractiveness of ICT careers has shifted substantially over a relatively short period.

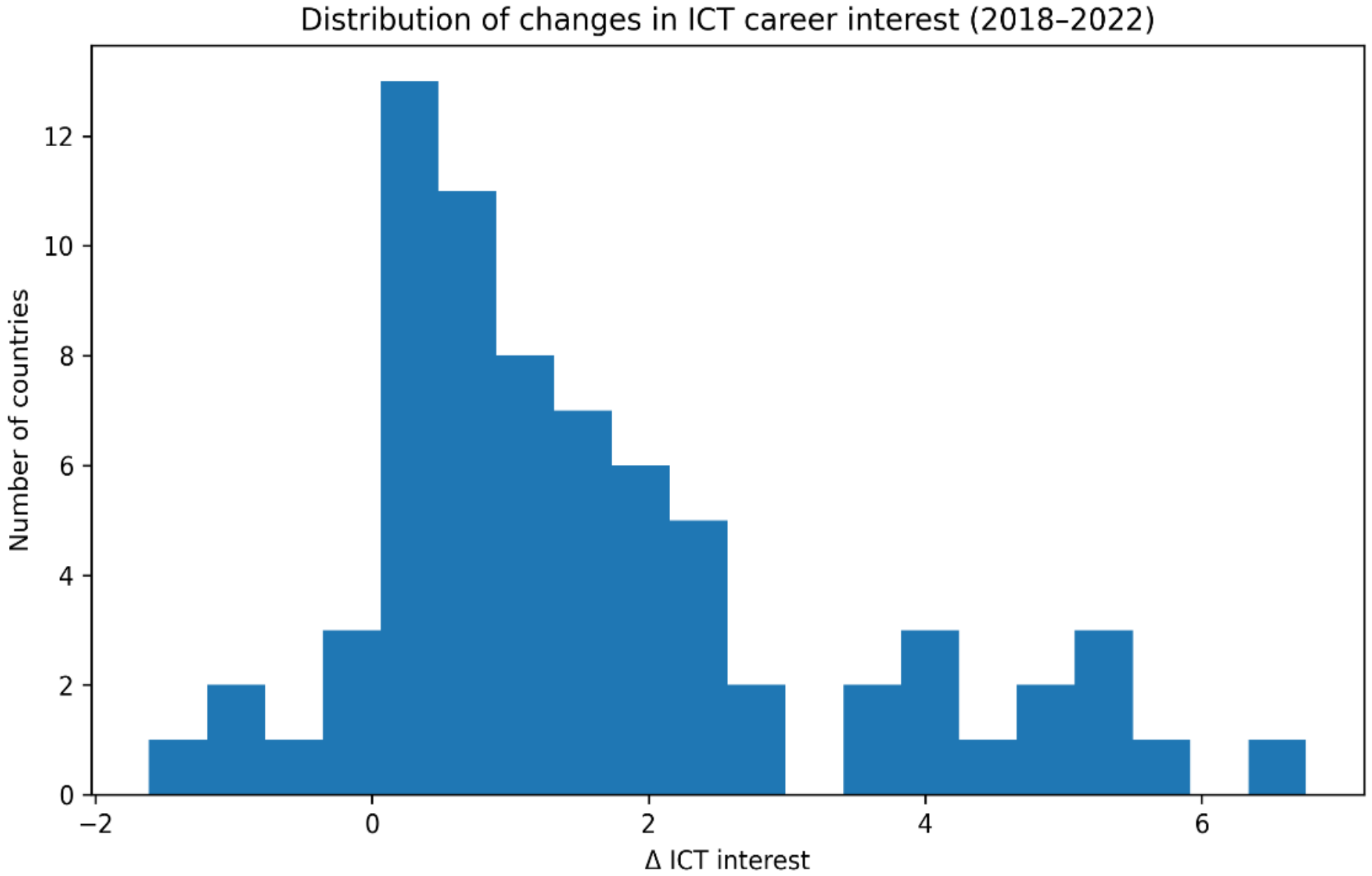

Figure 2. Distribution of ICT aspiration change (2018–2022)

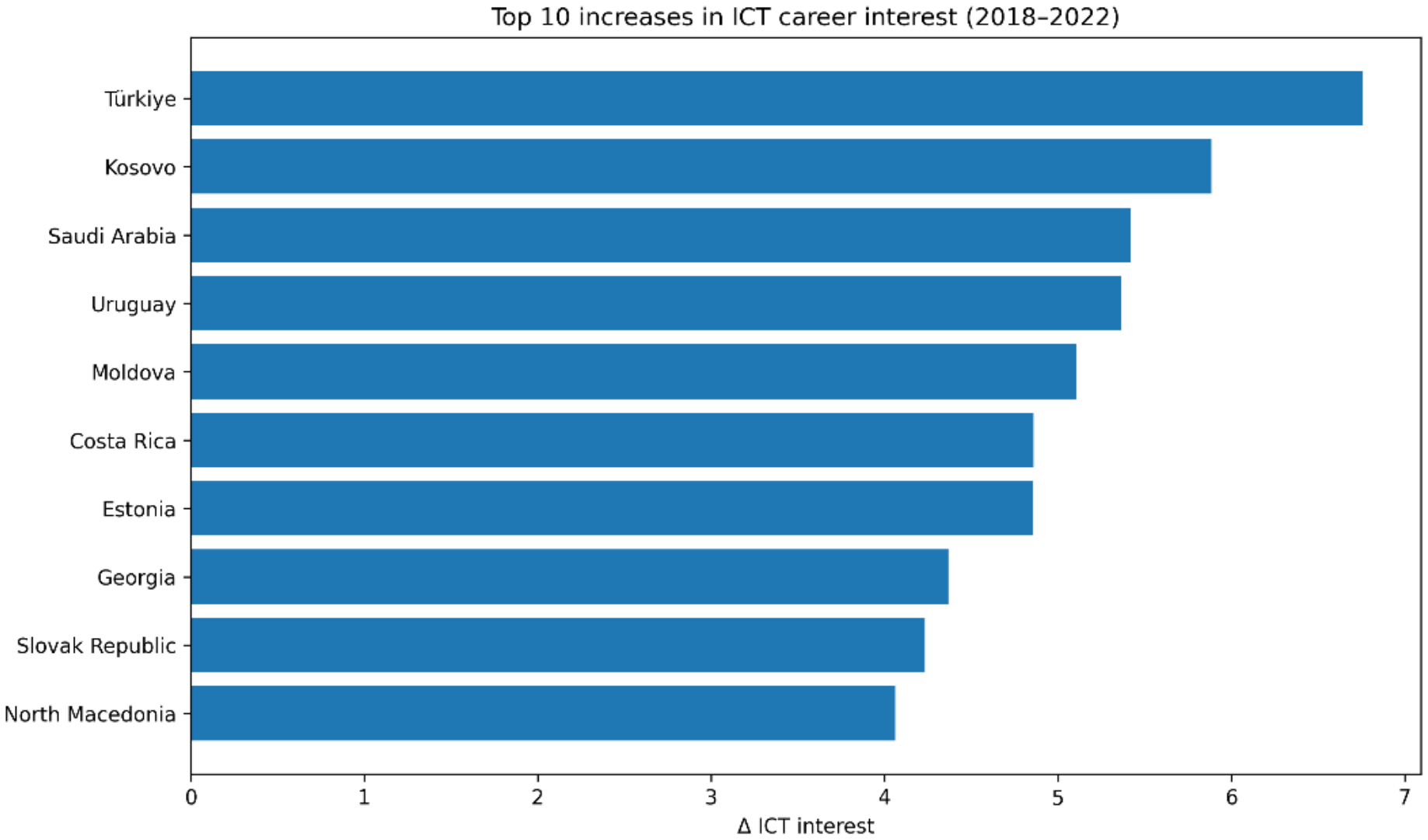

Figure 3. Top 10 increases in ICT career aspirations (2018–2022)

Figure 4 displays the ten countries with the largest declines in ICT career interest. The negative changes observed in countries such as Poland, the Netherlands and Baku (Azerbaijan) indicate that, despite global narratives emphasizing digitalization, ICT careers did not uniformly gain attractiveness. In some contexts, declining interest may reflect saturation effects in ICT labor markets, perceived barriers to entry or competition from alternative professional domains, highlighting that the diffusion of ICT-oriented aspirations is not linear or universal, but mediated by national education systems, labor market conditions and socio-cultural perceptions of technological careers.

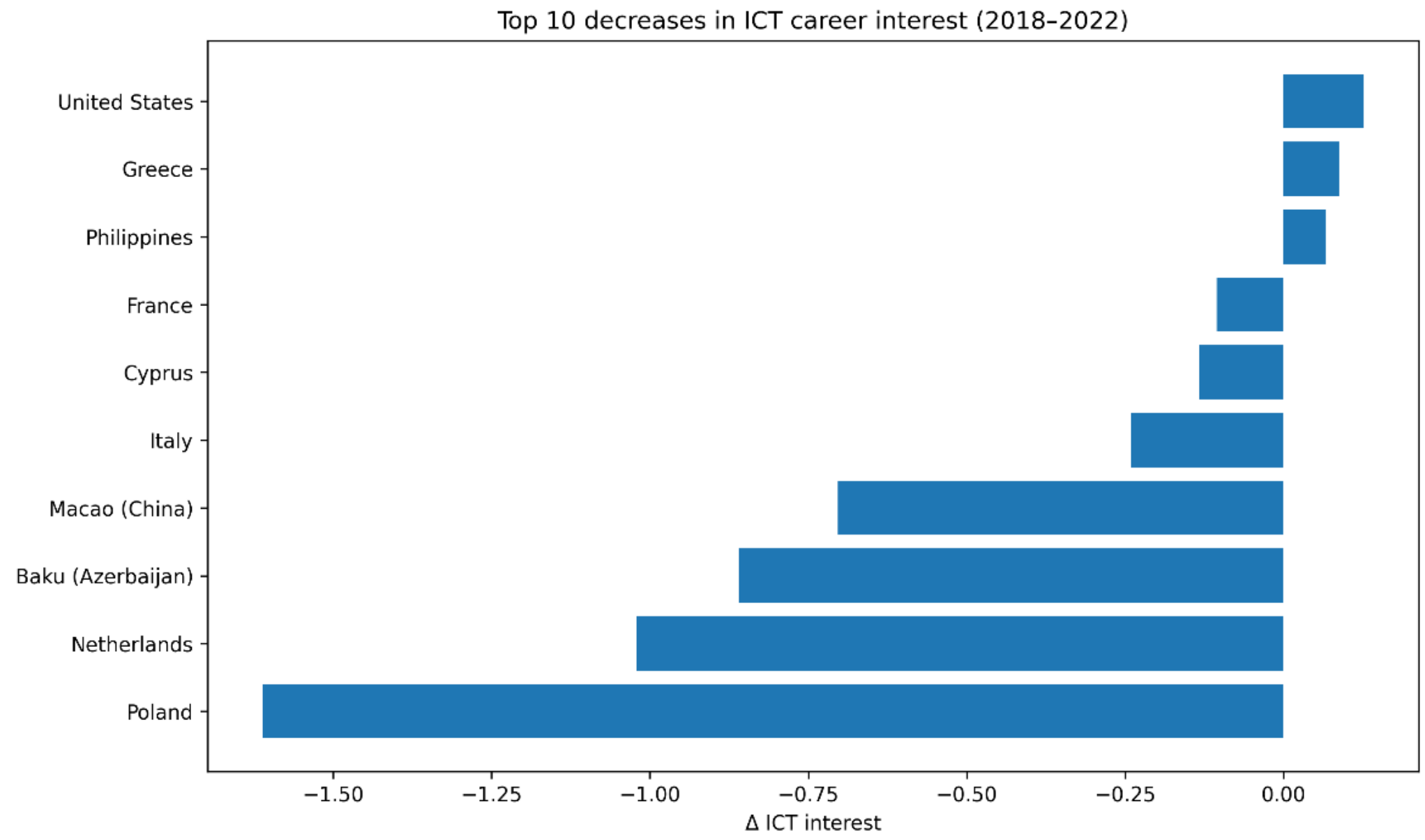

Figure 4. Top 10 decreases in ICT career aspirations (2018–2022)

Additionally, Figure 5 examines the association between changes in interest in Science & Engineering careers and changes in ICT career interest. The dispersion of points suggests a weak-to-moderate positive relationship, indicating that increases in ICT interest are often, but not always, accompanied by increases in broader STEM-oriented aspirations. Several countries display strong positive ICT growth despite stagnant or declining interest in traditional Science & Engineering fields, suggesting a partial decoupling between ICT and classical STEM trajectories. This pattern implies that ICT careers are increasingly perceived as a distinct and attractive pathway, not merely as a subset of general STEM professions.

Conversely, the presence of countries with declining Science & Engineering interest and modest ICT growth points to differentiated structural dynamics across scientific and technological domains.

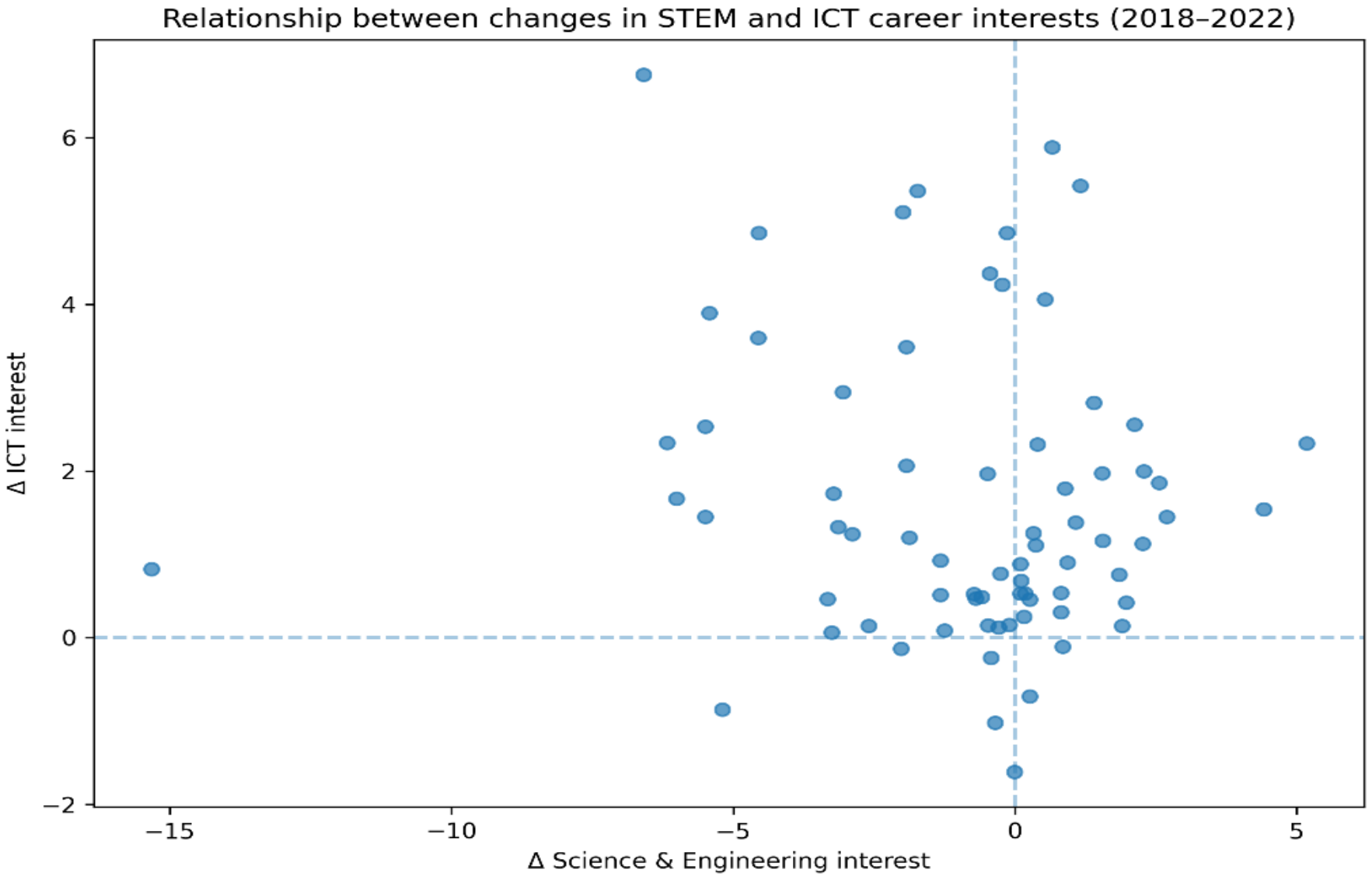

Figure 5. Relationship between changes in ICT and Science & Engineering aspirations (2018–2022)

The heatmap in Figure 6 provides a comparative overview of changes in career interest across four domains (Health, Science & Engineering, Science-related technical occupations and ICT) for the countries with the largest variation. The results reveal pronounced cross-domain heterogeneity, with several education systems showing strong increases in ICT interest alongside more moderate or even negative changes in traditional scientific and health-related careers. At the same time, some countries exhibit parallel increases across multiple domains, indicating more general expansions in career expectations rather than sector-specific shifts. ICT growth appears to coexist with divergent trajectories in other fields, reflecting both substitution and complementarity effects.

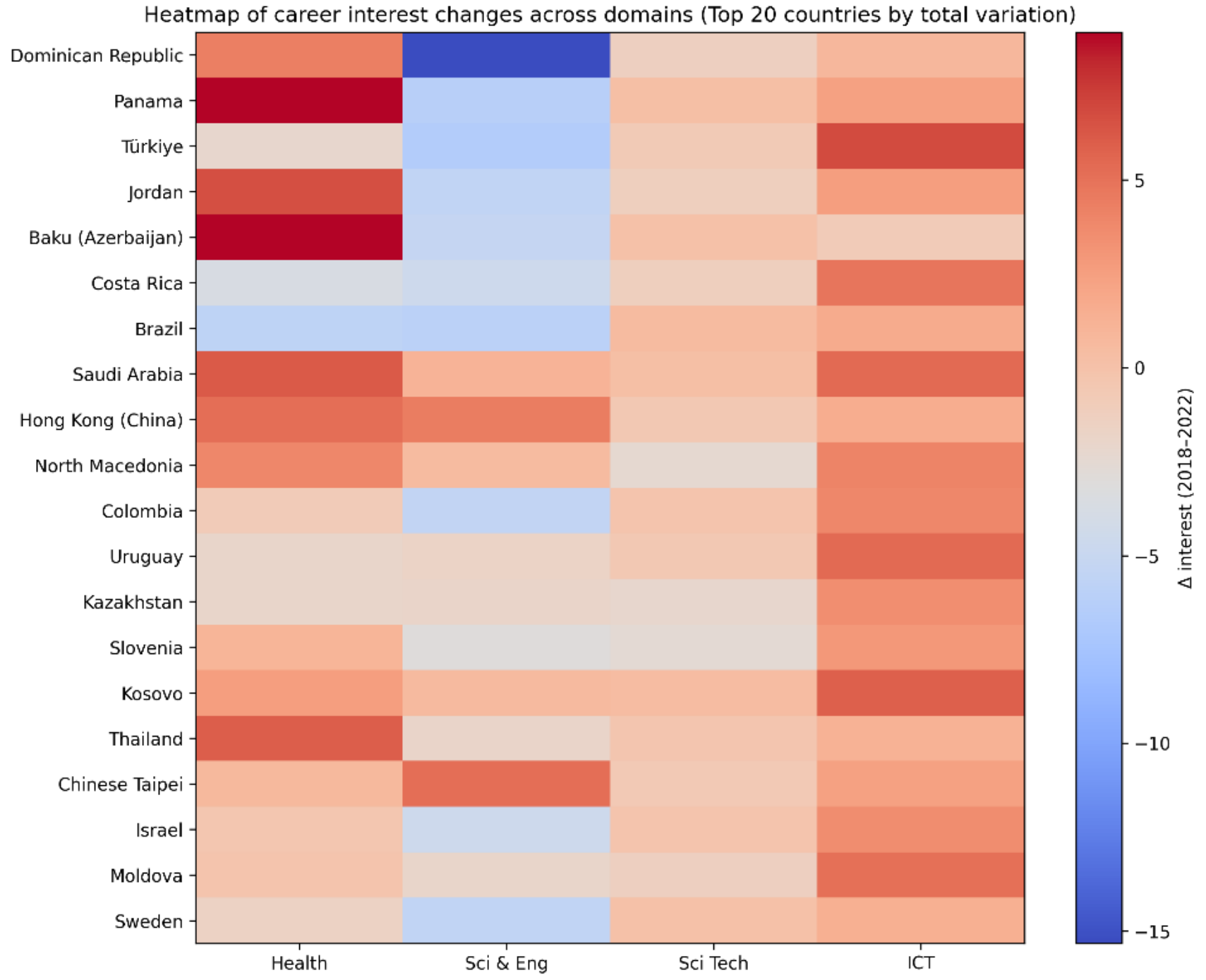

Figure 6. Heatmap of multidomain changes in career aspirations (Top 20 most volatile countries)

The distributional analysis in Figure 7 further reinforces these differences across domains. ICT exhibits a clearly positive median change and several high positive outliers, indicating consistent growth across countries. In contrast, Science & Engineering shows a slightly negative median and substantial dispersion, reflecting heterogeneous and, in some cases, declining interest. The Health domain presents a more balanced distribution around a small positive median, but with wide variability across countries. Science-related technical occupations display the lowest dispersion and a median close to zero, suggesting relatively stable career expectations, highlighting a structural divergence between ICT and the other domains. ICT stands out as the only field with a consistently positive shift, whereas traditional STEM and health-related careers exhibit more fragmented and context-dependent trajectories.

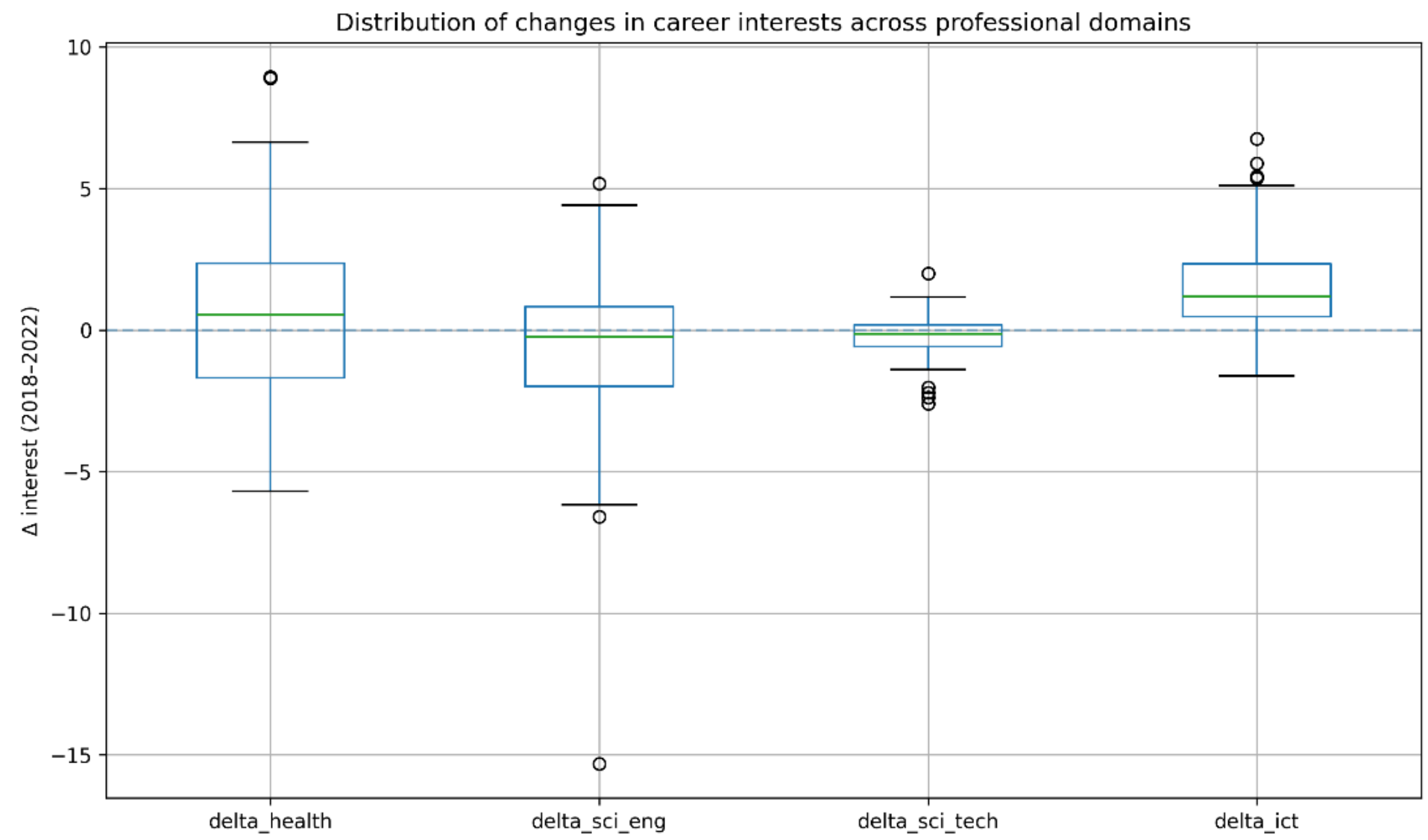

Figure 7. Boxplot of changes in career aspirations by domain (2018–2022)

These findings document a structurally uneven transformation of career aspirations toward ICT between 2018 and 2022. While the overall distribution indicates a global tendency toward increased ICT interest, the country-level extremes reveal substantial divergence in how education systems respond to digitalization. The weak coupling between ICT and traditional STEM changes suggests that ICT careers are becoming an autonomous aspiration domain, shaped not only by scientific education but also by the visibility of digital industries, platform economies and AI-driven labor markets.

**4.2. Latent readiness profiles and AI preparedness of education systems**

The second layer of results focuses on the latent profiling of education systems based on learning environment indicators, with the aim of uncovering structural patterns of readiness for digital and AI-related transformations. A VAE is employed to project standardized indicators of student autonomy, digital learning skills and teacher support into a low-dimensional latent space, which can be interpreted as a continuous representation of educational readiness. The two-dimensional latent representation derived from the VAE, based on standardized indicators of student autonomy, digital skills and teacher support, reveals substantial cross-national heterogeneity in structural readiness for digital and AI-related transformations. Countries are distributed across all quadrants of the latent space (Figure 8), without forming clearly separated clusters, indicating that educational readiness is not a categorical property but a multi-dimensional continuum shaped by different combinations of autonomy, digital competencies and institutional support. The absence of sharp boundaries suggests overlapping profiles of readiness, with gradual transitions between lower- and higher-capacity systems. Countries positioned in regions with higher latent values can be interpreted as having more favorable structural conditions for AI integration, while those located in lower regions reflect weaker foundational capacities. The results support the conceptualization of AI readiness as an emergent latent construct arising from the joint configuration of multiple learning environment dimensions, rather than from any single observable indicator.

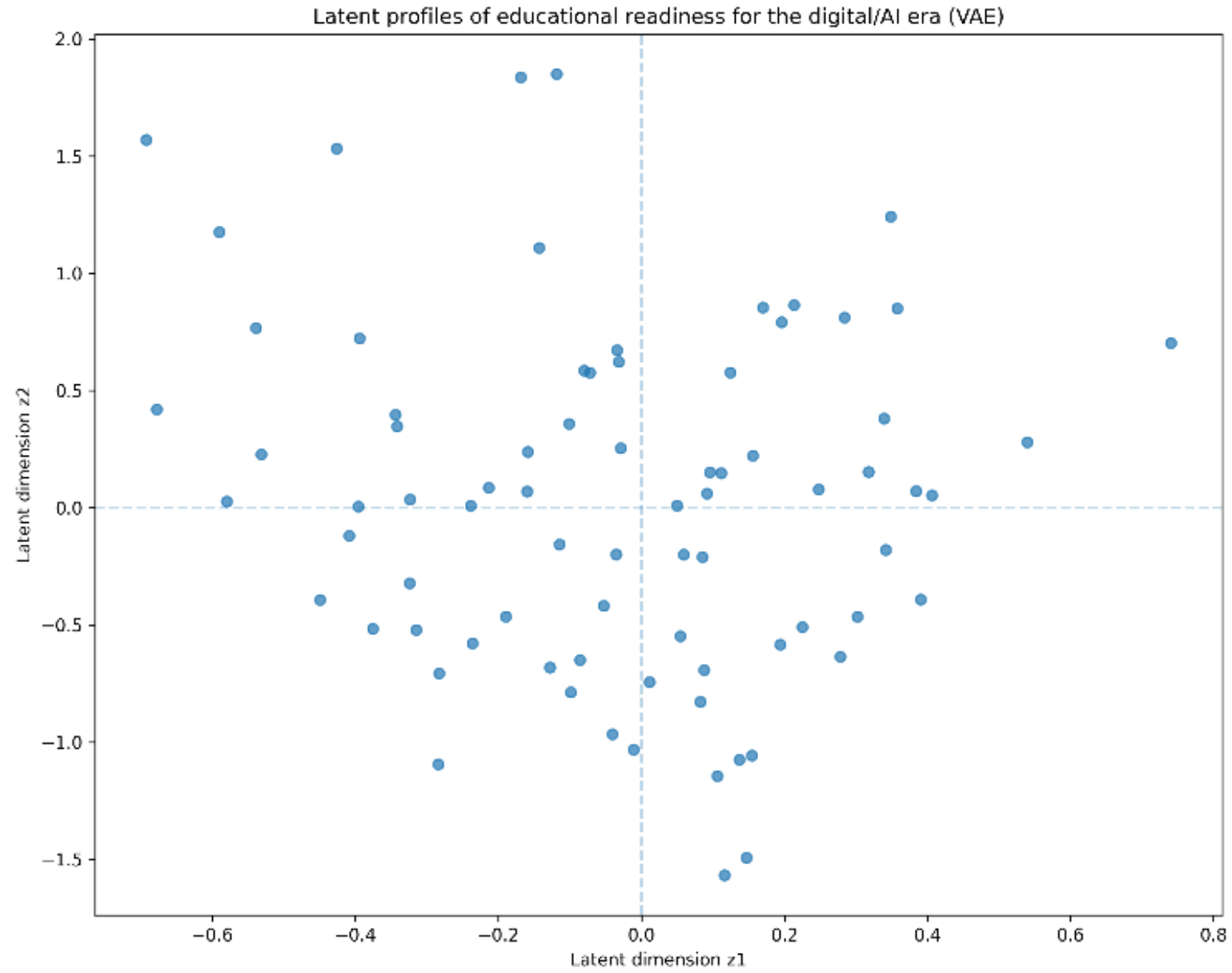

Figure 8. VAE latent space of education systems

The latent space representation is extended by encoding ICT career aspiration growth (2018–2022) as a color gradient, enabling an integrated assessment of the relationship between structural educational readiness and changes in ICT-related career interest (Figure 9). A clear, although imperfect, gradient emerges, countries with higher levels of latent readiness tend, on average, to exhibit stronger increases in ICT career aspirations. This pattern suggests that favorable learning environments, characterized by higher autonomy, stronger digital skills and greater teacher support, facilitate the translation of digital transformation into student career expectations. At the same time, variation within similar regions of the latent space indicates that educational readiness alone does not fully determine ICT aspiration growth. Countries with comparable structural profiles display heterogeneous outcomes, highlighting the role of additional contextual factors.

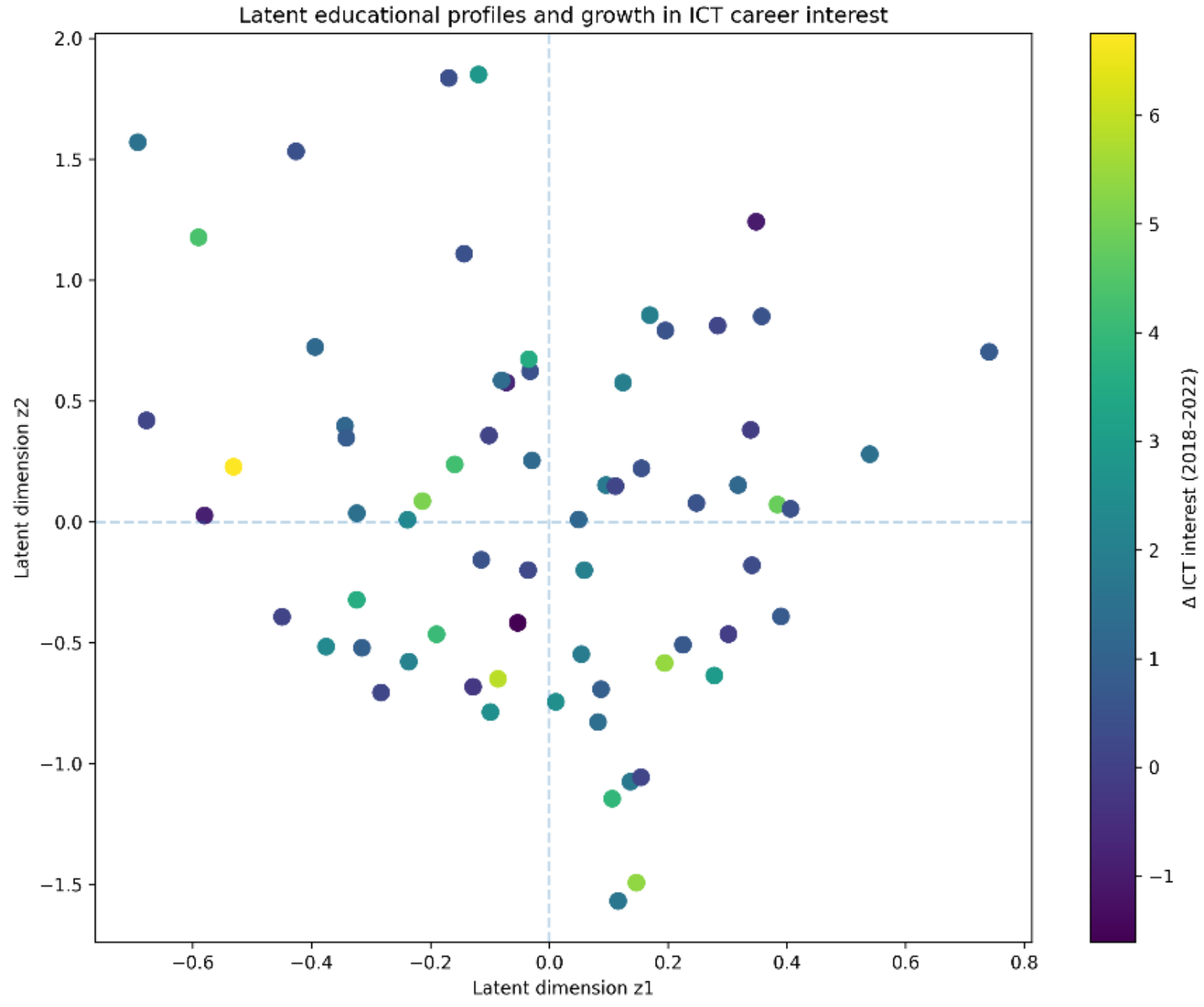

Figure 9. VAE latent space colored by ICT aspiration change (2018–2022)

The ranking of countries based on the composite AI Readiness Index reveals a strong concentration of high-readiness profiles among advanced education systems, with countries such as Korea, Finland, the Netherlands and New Zealand occupying the leading positions (Figure 10). These systems are characterized by higher levels of student autonomy, strong digital competencies and robust teacher support, reflecting their capacity to integrate AI-related practices into educational processes. The presence of both European and non-European countries among the top performers indicates that AI readiness is not region-specific but emerges from diverse institutional pathways.

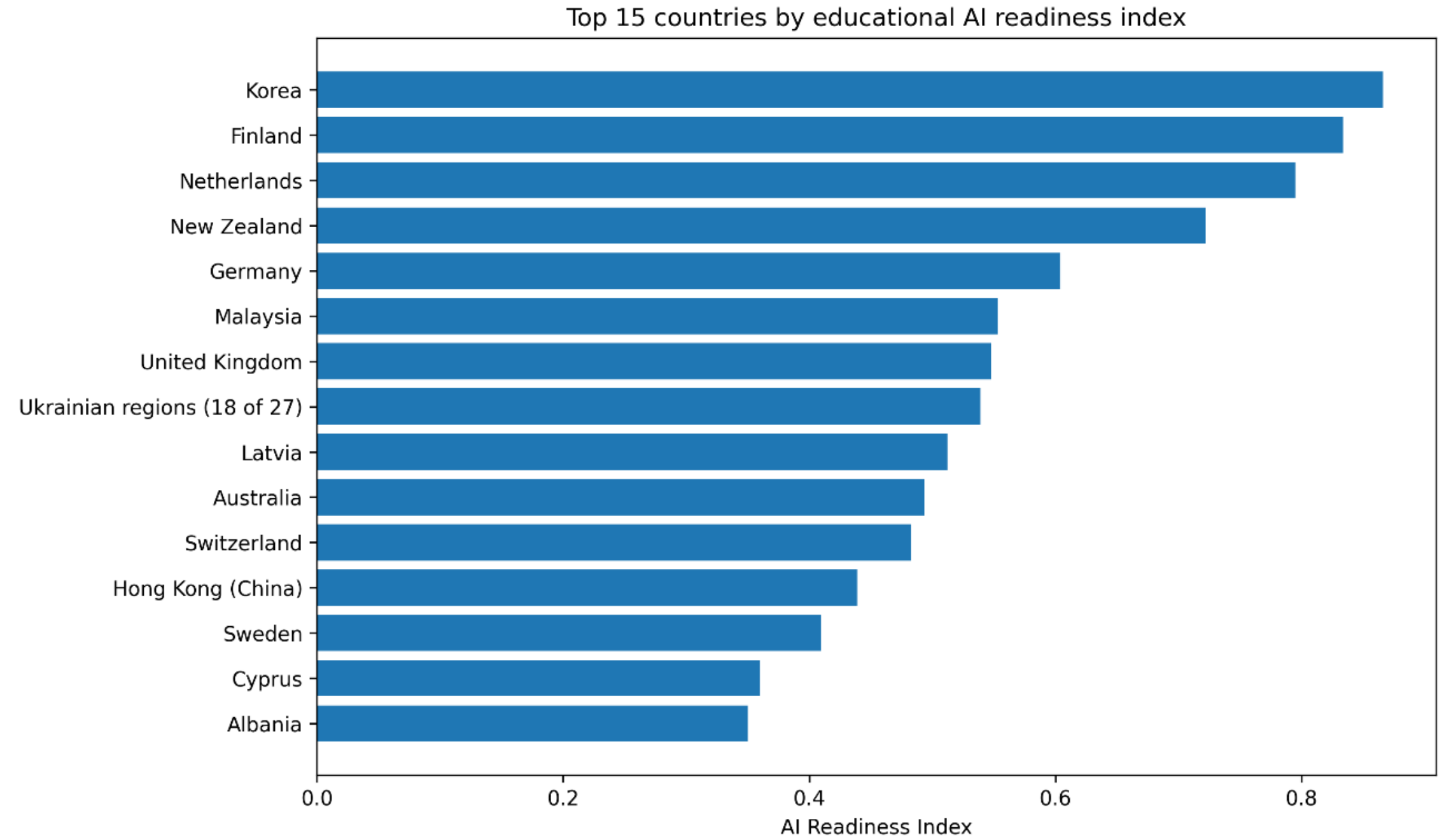

Figure 10. Top 15 education systems by AI readiness index

**4.3. Learning environment effects, typologies and cross-domain consistency**

Systems combining high autonomy and strong digital skills represent more favorable configurations, where institutional flexibility is associated with higher digital competence. In contrast, systems characterized by both low autonomy and weak digital skills reflect more constrained environments and lower digital capacity.

The joint distribution of school autonomy, digital skills and teacher support highlights the multi-dimensional nature of educational readiness for digital transformation. Education systems characterized by high autonomy and strong digital skills, combined with substantial teacher support, form structurally favorable environments for the effective integration of digital tools (Figure 11). In contrast, systems with low autonomy and weak digital skills reflect compounded structural disadvantages. At the same time, some countries exhibit strong digital skills alongside limited teacher support, indicating potential sustainability constraints (Figure 12).

Intermediate clusters in Figure 13 indicate that autonomy alone is not sufficient to ensure strong digital outcomes. Some systems exhibit relatively high autonomy but only moderate digital performance, while others achieve reasonable digital skills despite limited autonomy. The typology highlights substantial heterogeneity across education systems and suggests that the relationship between autonomy and digital skills is conditional on broader systemic factors.

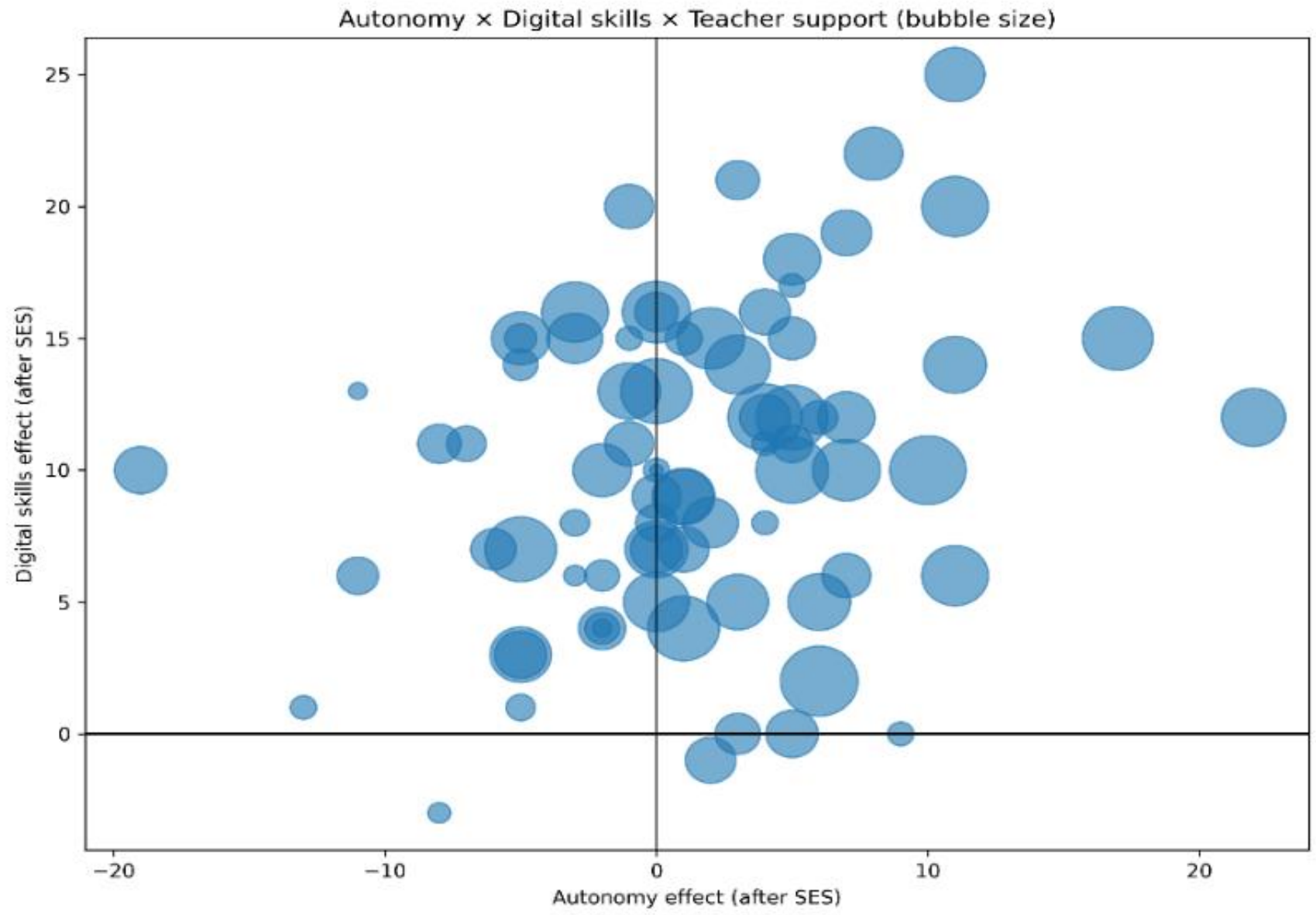

Figure 11. Autonomy × digital skills × teacher support (bubble plot, mathematics)

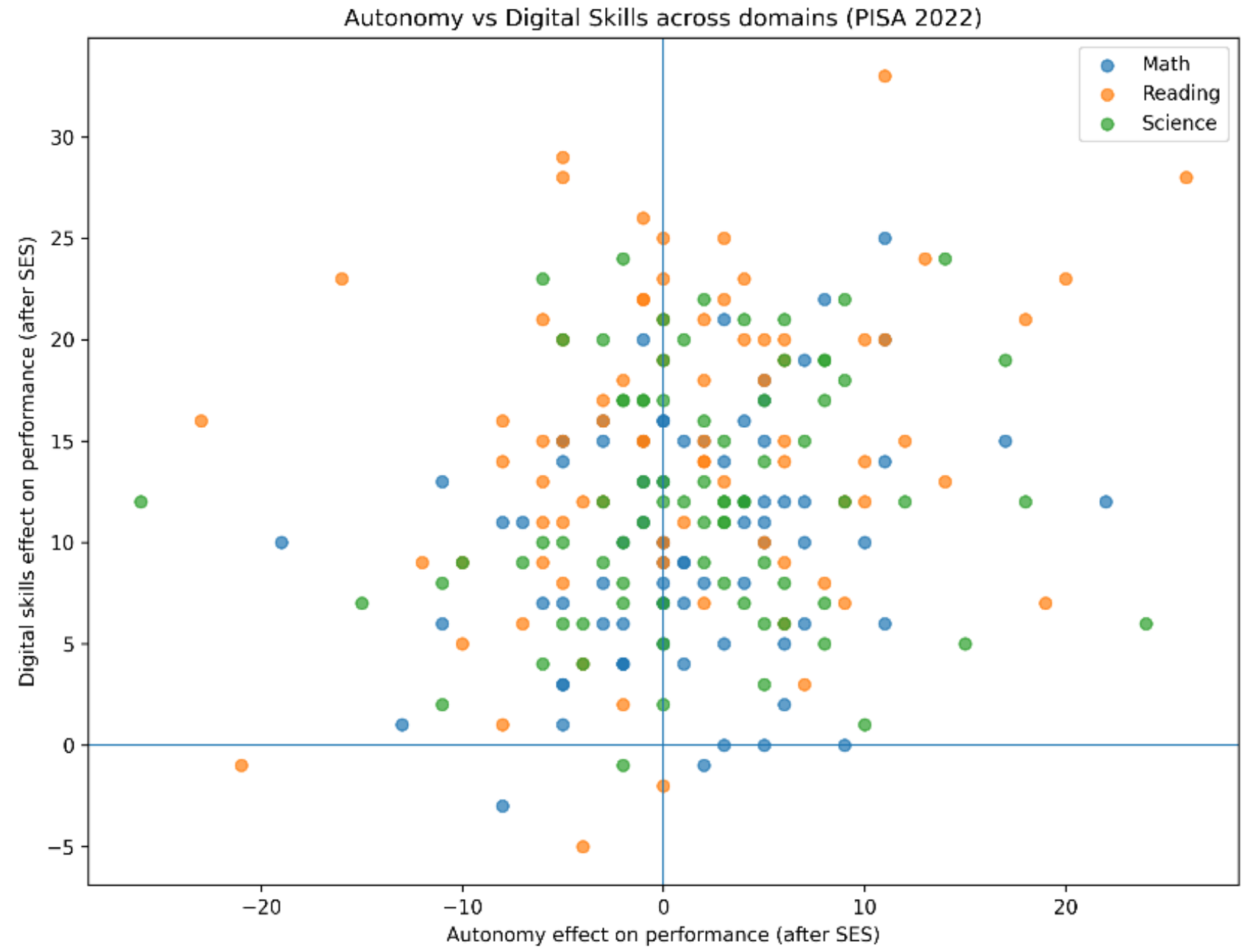

Figure 12. Autonomy versus digital skills effects on performance (PISA 2022)

The relationship between school autonomy and digital skills across mathematics, reading and science is weakly positive but highly dispersed, indicating substantial cross-national heterogeneity (Figure 14). While some education systems exhibit a clear positive association, where higher autonomy coincides with stronger digital competencies, many countries deviate from this pattern. The dispersion varies across domains, with reading and science showing a higher concentration of observations in the upper range compared to mathematics. The wide spread of observations suggests that autonomy is not a universal driver of digital skills, but rather an enabling condition whose effectiveness depends on broader systemic factors.

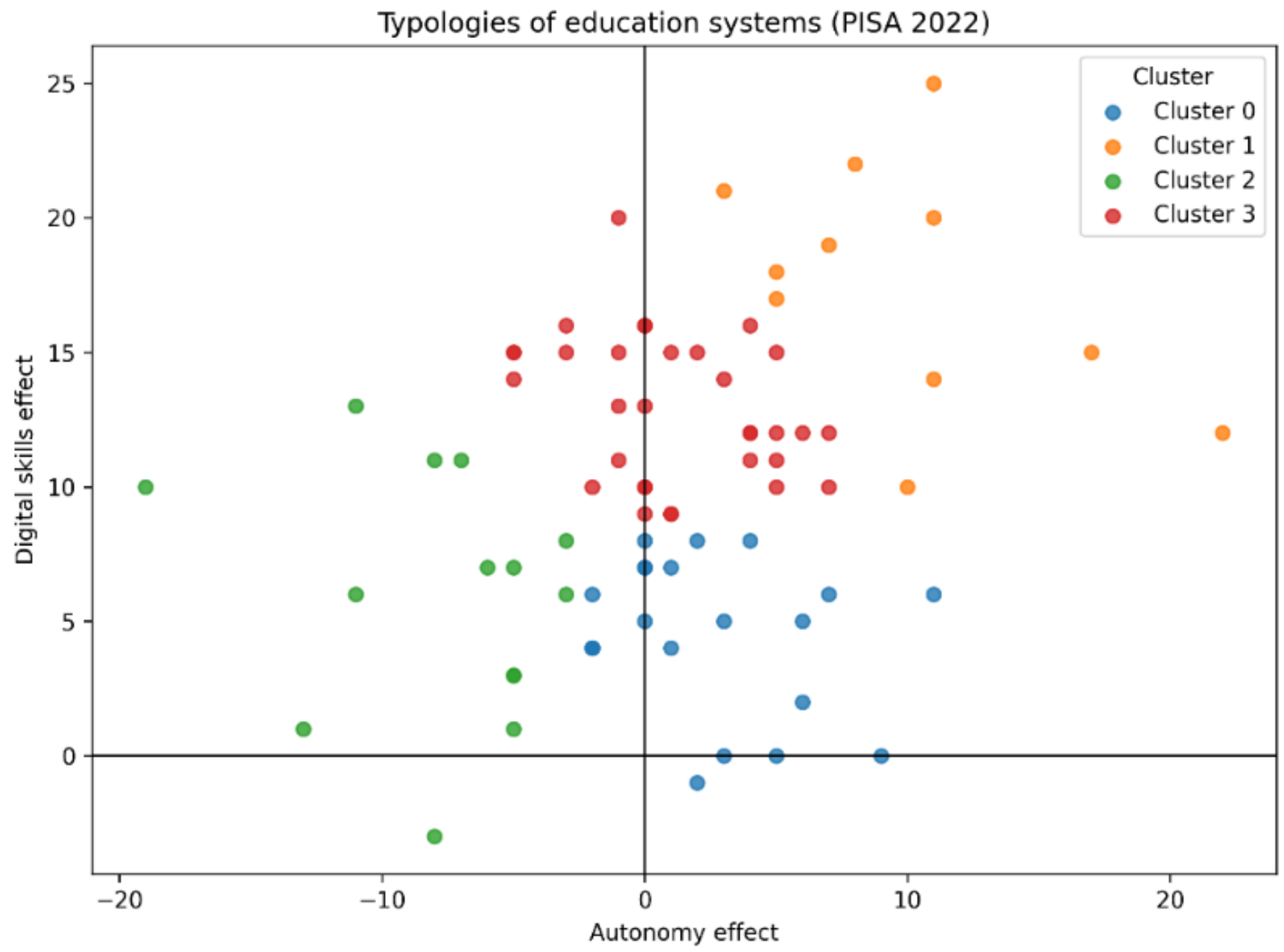

Figure 13. Clusters of education systems based on autonomy and digital skills (mathematics)

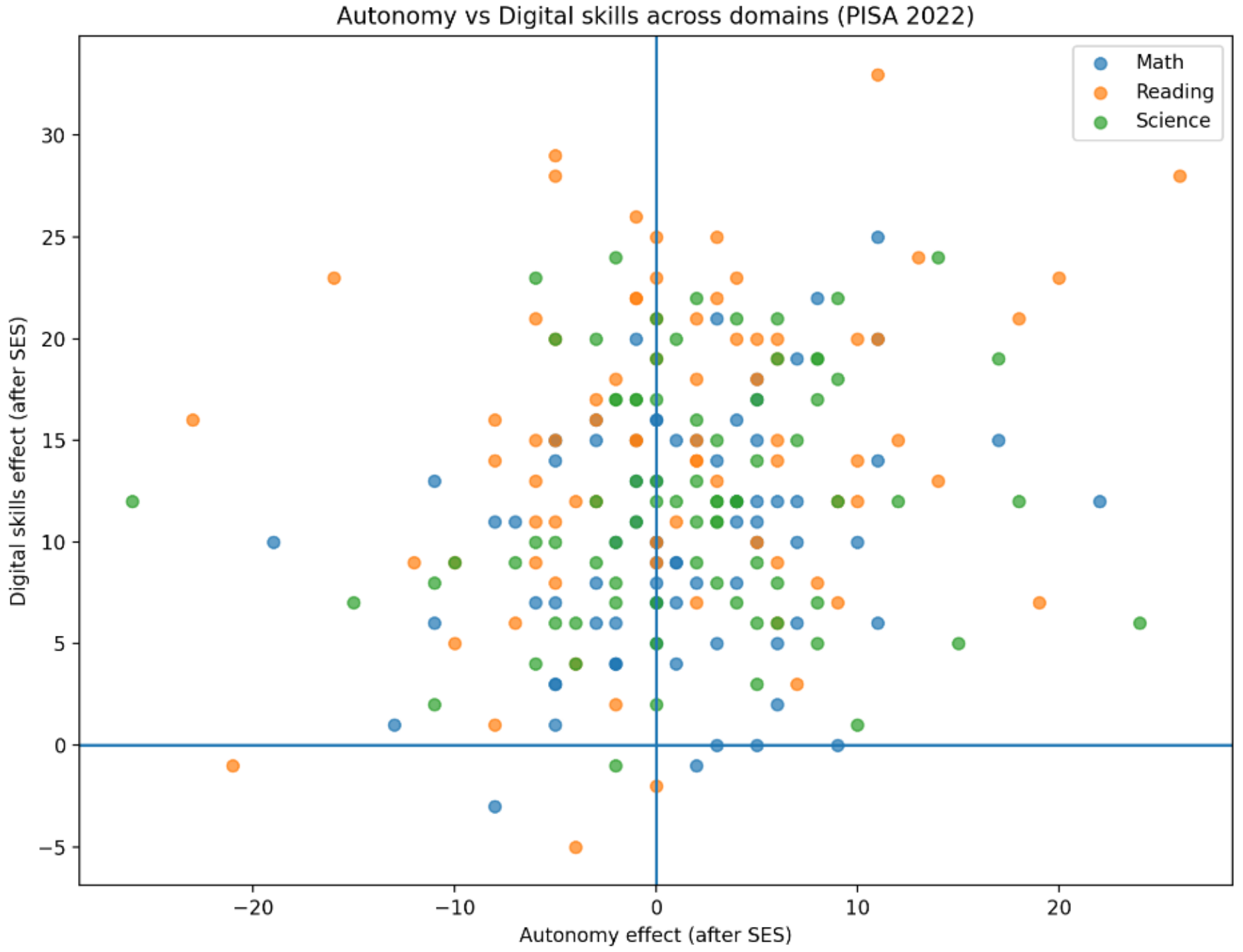

Figure14. Autonomy versus digital skills effects across domains (mathematics, reading, science)

**4.4 Distribution and heterogeneity of ICT career aspiration change (2018–2022)**

The distribution of changes in ICT career aspirations between 2018 and 2022 is positively skewed, with most countries exhibiting moderate increases, typically between 0 and 2 percentage points (Figure 15). At the same time, the presence of a right tail indicates that a small number of countries experienced substantially larger increases, while a limited number of systems show stagnation or declines, highlighting significant cross-national heterogeneity in ICT aspiration dynamics. The asymmetry indicates that the drivers of ICT aspiration growth are context-dependent, with some countries experiencing stronger shifts than others.

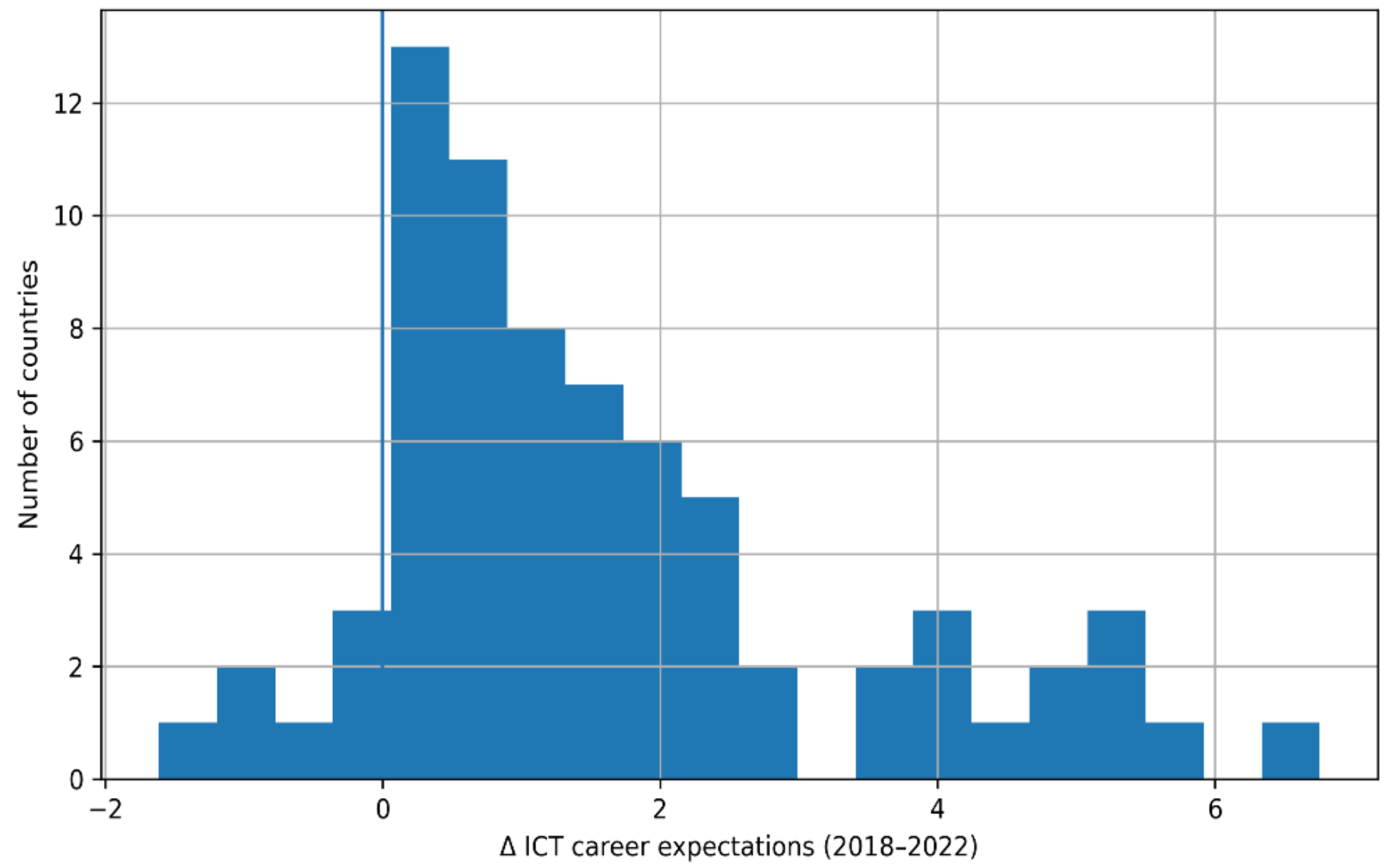

Figure 15. Distribution of ICT career aspiration change (PISA 2018–2022)

**4.5 Interaction between school autonomy and teacher support in shaping digital skills**

The interaction between school autonomy and teacher support reveals a conditional relationship in the development of digital skills (Figure 16). In contexts with lower levels of teacher support, the effect of autonomy is stronger, suggesting that autonomy may act as a compensatory mechanism by enabling schools to adapt learning practices in the absence of strong institutional guidance. Conversely, in systems with high levels of teacher support, the marginal effect of autonomy is reduced, as existing support structures already facilitate the effective development of digital skills,

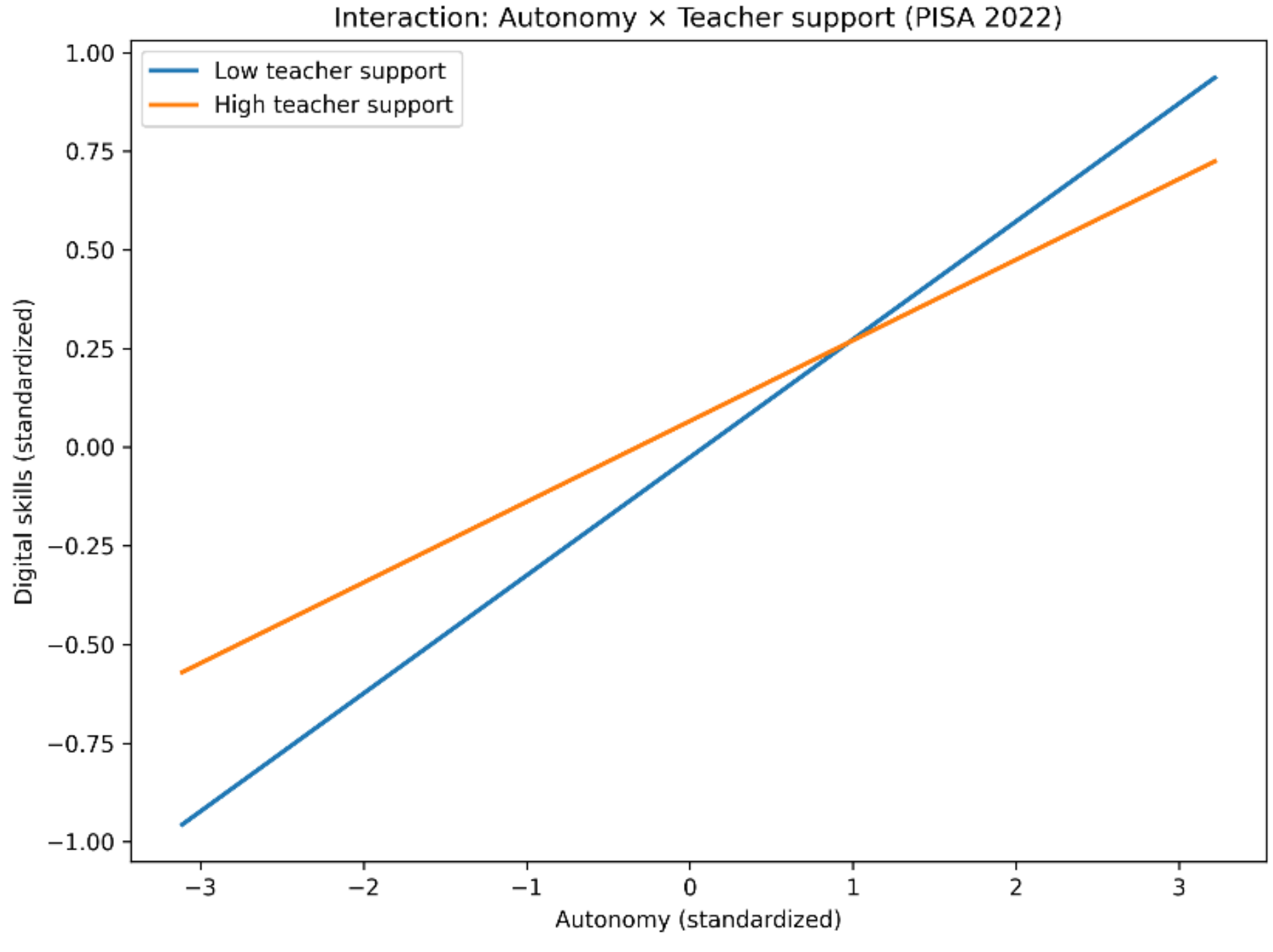

Figure 16. Interaction: Autonomy × Teacher support (PISA 2022)

**4.6 Autonomy and digital skills in mathematics (domain-specific association)**

The interaction between school autonomy and teacher support reveals a conditional relationship in the development of digital skills. In contexts with lower levels of teacher support, the effect of autonomy is stronger, suggesting that autonomy may act as a compensatory mechanism by enabling schools to adapt learning practices in the absence of strong institutional guidance. Conversely, in systems with higher levels of teacher support, the marginal effect of autonomy is reduced, as existing support structures already facilitate the effective development of digital skills.

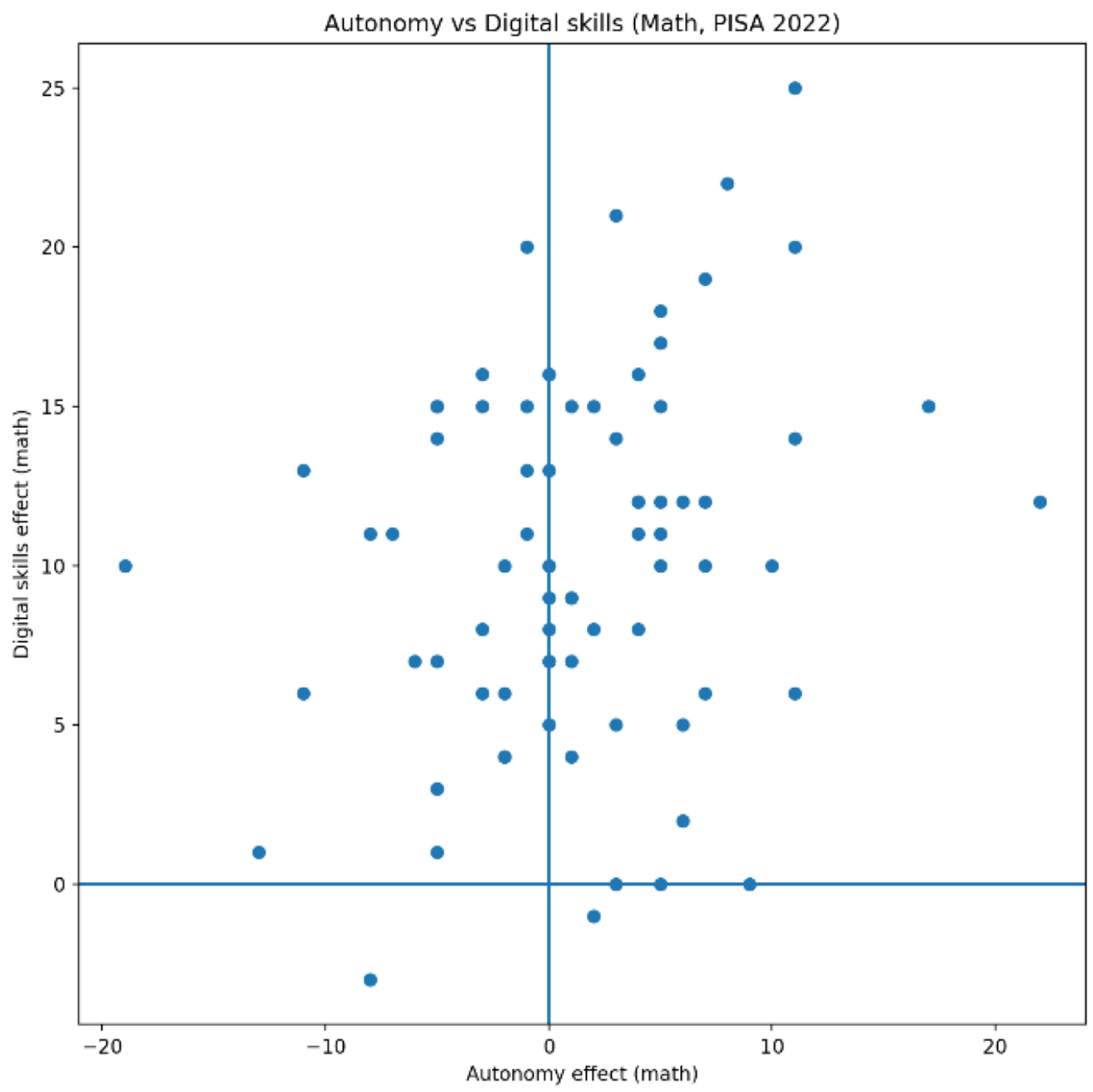

Figure 17. Autonomy vs. digital skills (Math, PISA 2022)

### 4.7. Gender disparities in autonomy, digital learning and ICT-related outcomes

The relationship between gender differences in perceived autonomy and the effect of autonomy on mathematics performance is highly dispersed, indicating no systematic alignment between the two (Figure 18).

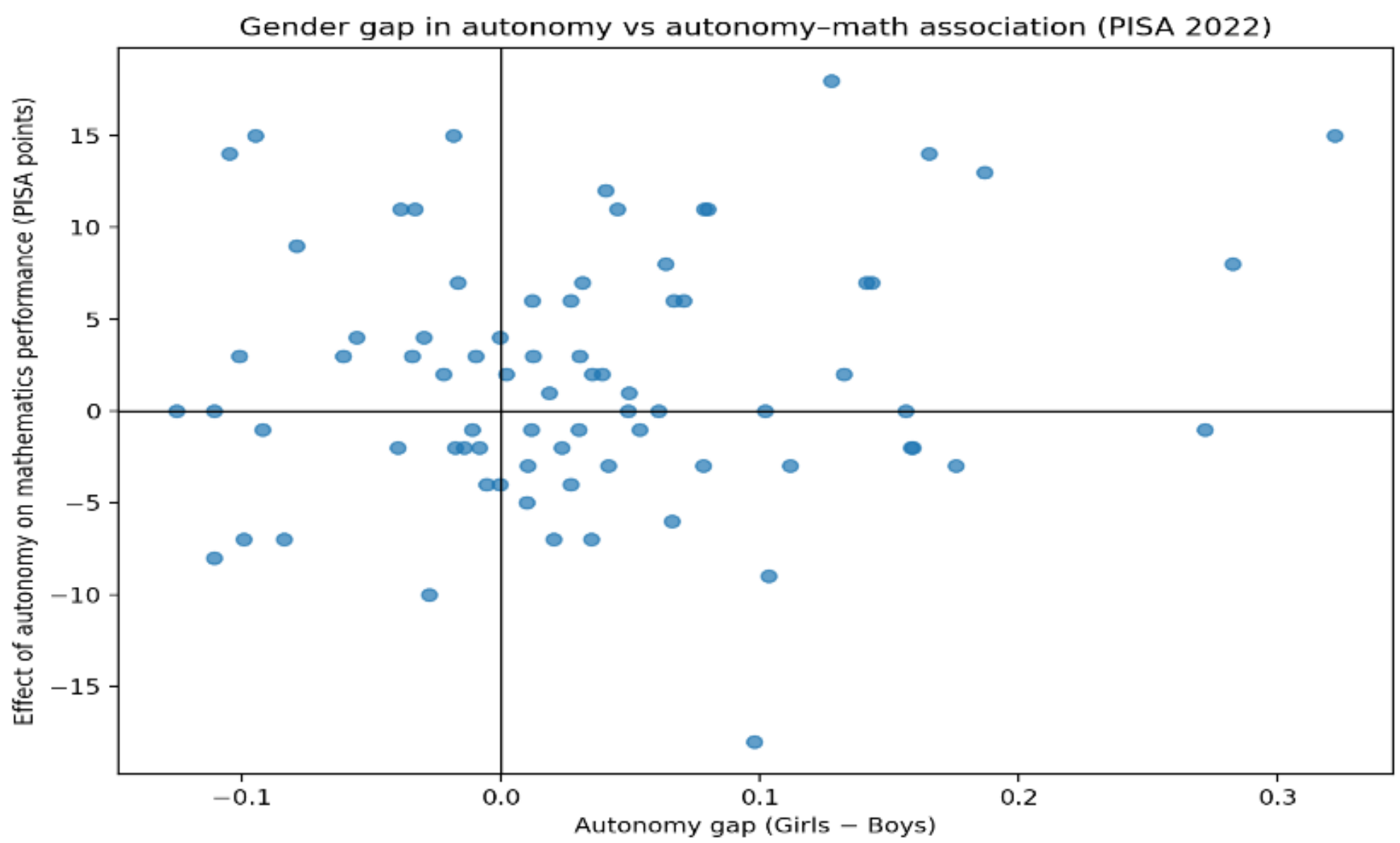

Figure 18. Gender differences in autonomy effects

Across countries, positive or negative autonomy effects coexist with both female and male advantages in perceived autonomy, suggesting that gender differences in autonomy are not consistently associated with performance outcomes. This pattern implies that the effectiveness of autonomy is largely gender-neutral in aggregate terms, while gendered experiences of autonomy vary substantially across education systems. The absence of a clear relationship indicates that differences in access to autonomy do not automatically translate into differential academic returns.

**4.8. Linear discriminant analysis of education systems by ICT aspiration growth**

This section examines whether structural characteristics of education systems can discriminate between countries with high versus low growth in ICT career aspirations. A Linear Discriminant Analysis (LDA) model was estimated using standardized indicators of school autonomy, digital skills and teacher support, with countries classified based on their level of ICT aspiration growth. The model exhibits moderate classification performance, indicating that learning environment indicators have meaningful but limited discriminative power, suggesting that, while these factors are relevant, they do not fully account for differences in ICT career dynamics (Figure 19). The estimated coefficients show that digital skills have the strongest positive association with high ICT aspiration growth, followed by teacher support. In contrast, the contribution of autonomy is comparatively weaker, reinforcing the idea that autonomy alone is insufficient in the absence of complementary instructional support. The LDA model was evaluated using stratified cross-validation in order to assess its generalization performance. The average classification accuracy obtained across folds indicates that the structural learning environment indicators possess non-trivial discriminative power in separating countries by ICT aspiration growth trajectories. However, the accuracy remains moderate rather than high, suggesting that while these features contain relevant information, they do not fully determine the differences observed in ICT career dynamics.

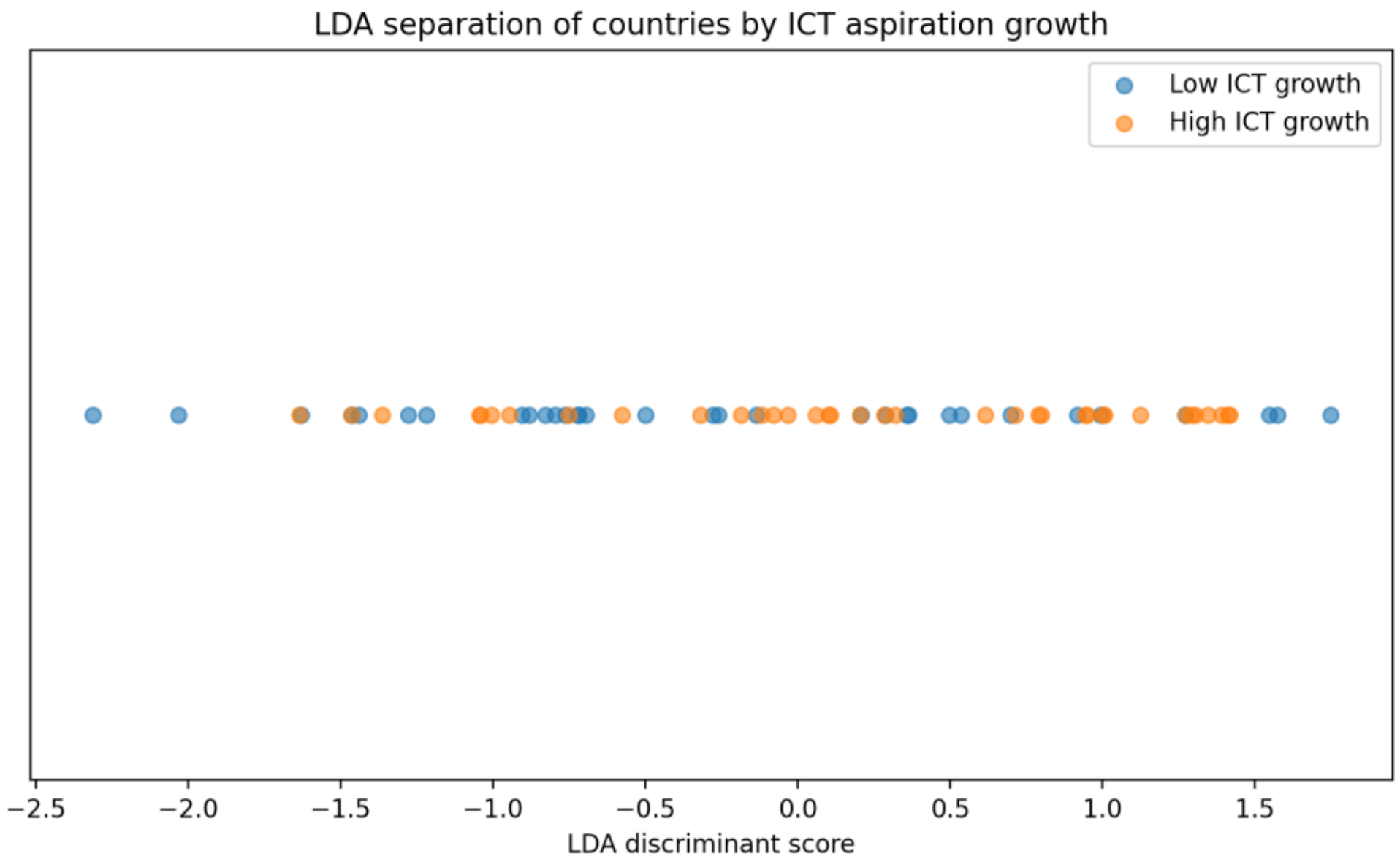

Figure 19. Linear Discriminant Analysis of education systems by ICT aspiration growth

Countries with higher ICT aspiration growth tend to exhibit higher discriminant scores, while low-growth systems are concentrated in the lower range. However, the substantial overlap between groups indicates that no clear boundary separates growth trajectories based solely on educational indicators. This overlap highlights that ICT career dynamics are not fully determined by school-level conditions. Instead, learning environments act as enabling factors that increase the likelihood of ICT-oriented aspirations, without guaranteeing their emergence.

**4.9. Regression results**

The regression results indicate that teacher support and autonomy exhibit negative associations with ICT career aspiration growth, while digital skills show a small positive effect. The standardized coefficients reveal that teacher support has the strongest (negative) association ($\beta = -0.324$), followed by autonomy ($\beta = -0.192$), whereas digital skills have a weaker positive effect ($\beta = 0.072$). This pattern suggests that higher levels of digital skills are associated with increased ICT career aspirations, confirming their role as a key enabling factor. In contrast, the negative coefficients for autonomy and teacher support indicate that their relationship with ICT aspiration growth is not straightforward and may reflect more complex or context-dependent dynamics.

One possible interpretation is that, in some education systems, higher levels of institutional support and structured learning environments may be associated with more traditional academic pathways, while ICT-oriented aspirations emerge more strongly in contexts characterized by different structural or labor market conditions. The regression results highlight that digital skills represent the only consistently positive predictor of ICT career aspiration growth, while autonomy and teacher support exhibit weaker and non-linear relationships.

**4.10. Counterfactual simulation**

The counterfactual simulation evaluates the effect of increasing school autonomy by one standard deviation while holding digital skills and teacher support constant. The results indicate that the predicted increase in ICT career aspirations is modest on average and exhibits substantial variation across countries. In most cases, the simulated increase in autonomy leads to only limited changes in predicted ICT aspiration levels, suggesting that, although autonomy contributes to ICT career orientation, its marginal impact is relatively small compared to the effects associated with digital skills and teacher support.

At the same time, the comparison between observed and predicted values reveals that the model does not fully capture extreme cases of ICT aspiration growth. Several countries with high observed increases in ICT interest display much lower predicted values under the counterfactual scenario, indicating that additional contextual factors beyond learning environments influence these dynamics. The counterfactual analysis confirms that policy interventions focused exclusively on increasing autonomy are unlikely to generate substantial shifts in ICT career aspirations in the absence of parallel improvements in digital competencies and pedagogical support.

**4.11 Probabilistic relationships between career domains**

The Bayesian network analysis does not identify significant conditional dependencies between changes in ICT career aspirations and shifts in other occupational domains. The learned structure contains no edges, indicating that changes in ICT aspirations are not systematically linked to contemporaneous changes in health, science and engineering, or science-related technical career expectations. The marginal distribution of ICT aspiration changes is evenly distributed across the three categories (low, medium, high), further suggesting the absence of dominant dependency patterns within the observed data, indicating that ICT career dynamics evolve relatively independently from other occupational domains, rather than emerging as a direct substitution or extension of traditional STEM pathways.

The findings support the interpretation that ICT career aspirations represent a distinct and increasingly autonomous domain of student expectations.

## 5. Conclusions

Our study investigated the relationship between learning environments and changes in students' ICT-related career aspirations in the context of digital and AI transformation. By combining harmonized PISA 2018–2022 data with latent representation learning and multivariate modeling, the analysis provides an integrated perspective on how structural features of education systems shape digital career trajectories.

The empirical results indicate a clear but uneven expansion of ICT career aspirations across countries. While most education systems experienced positive growth, the magnitude and direction of change vary substantially, highlighting significant cross-national heterogeneity. This suggests that exposure to digital transformation alone is insufficient to generate uniform shifts in students' occupational expectations.

Across all analytical approaches, digital skills emerge as the most consistent and influential factor associated with ICT aspiration growth, underscoring the importance of perceived digital competence as a key mechanism through which students evaluate the accessibility and attractiveness of ICT-related careers.

Teacher support plays a complementary role, reinforcing the relationship between digital skills and career expectations, indicating that supportive pedagogical environments enable the effective translation of digital exposure into sustained interest in ICT careers, highlighting the importance of instructional quality and teacher engagement in shaping the outcomes of digital learning processes.

By contrast, the role of autonomy appears more limited and strongly context-dependent. While autonomy can foster exploratory learning and engagement, its effects are contingent on the presence of adequate pedagogical support structures. In the absence of such support, autonomy alone does not appear sufficient to drive ICT-oriented career aspirations, challenging simplistic assumptions regarding the universal benefits of decentralization and emphasize the need for balanced and coordinated educational policies.

The latent representation analysis further reveals that educational readiness for the digital and AI era is best understood as a continuous and multi-dimensional construct. Education systems do not form discrete categories of "high" or "low" readiness but rather occupy positions along a continuum shaped by different combinations of autonomy, digital skills and teacher support. Systems characterized by favorable configurations across these dimensions tend to exhibit stronger ICT aspiration growth, although substantial variation persists within similar structural profiles.

Importantly, the findings also demonstrate that educational factors alone do not fully determine ICT career dynamics. Countries with comparable levels of structural readiness often display different outcomes, indicating that broader contextual influences, such as labor market conditions, national digital strategies and socio-cultural perceptions of ICT professions, play a significant role, reinforcing the interpretation of learning environments as enabling conditions rather than deterministic drivers.

From a policy perspective, the results highlight that investments in digital infrastructure are unlikely to produce sustained increases in ICT career interest in the absence of complementary pedagogical and institutional support. Effective educational strategies should therefore prioritize the development of digital competencies, the strengthening of teacher support systems and the integration of digital learning within coherent curricular frameworks.

**Funding**-This work was supported by a grant of the Ministry of Research, Innovation and Digitization, CNCS/CCCDI - UEFISCDI, project number COFUND-DUT-OPEN4CEC-1, within PNCDI IV.

**Acknowledgement**-This work was supported by a grant of the Ministry of Research, Innovation and Digitization, CNCS/CCCDI-UEFISCDI, project number COFUND-DUT-OPEN4CEC-1, within PNCDI IV.

**Conflicts of interest/Competing interests**-The authors declare that there is no conflict of interest.

**Ethics approval**-Not applicable

**Consent to participate**-Not applicable

**Consent for publication**-Not applicable

**Availability of data and material**-Data will be available on request.

**Authors' contributions**-D.M.P., A.B.: Methodology, Validation, Formal analysis, Investigation, Resources, Data Curation, Writing–Original Draft, Writing–Review and Editing, Visualization, Supervision. D.M.P., S.V.O.: Conceptualization, Formal analysis, Investigation, Writing–Original Draft, Writing–Review and Editing, Visualization, Project administration.